\documentclass[smallcondensed]{svjour3}
\usepackage{natbib}
\usepackage{graphicx,amsmath,amssymb}
\bibpunct{(}{)}{;}{a}{}{,}

\begin{document}

\title{Periodic Orbits and Escapes in Dynamical Systems}

\author{George Contopoulos \and Mirella Harsoula
\and Georgios Lukes-Gerakopoulos}

\institute{G. Contopoulos \at
Research Center for Astronomy and Applied Mathematics,
Academy of Athens,\\ Soranou Efesiou 4, GR-11527 Athens, Greece
\\\email{gcontop@academyofathens.gr}
\and M. Harsoula \at
Research Center for Astronomy and Applied Mathematics,
Academy of Athens,\\ Soranou Efesiou 4, GR-11527 Athens, Greece
\\\email{mharsoul@academyofathens.gr}
\and G. Lukes-Gerakopoulos \at
Research Center for Astronomy and Applied Mathematics,
Academy of Athens,\\ Soranou Efesiou 4, GR-11527 Athens, Greece \at
Theoretical Physics Institute, University of Jena, 07743 Jena, Germany
\\\email{gglukes@gmail.com}
}

\maketitle

\begin{abstract}
 We study the periodic orbits and the escapes in two different dynamical systems,
  namely (1) a classical system of two
 coupled oscillators, and (2) the Manko-Novikov metric (1992) which is a
 perturbation of the Kerr metric (a general relativistic system). We find their
 simple periodic orbits, their characteristics and their stability. Then we find
 their ordered and chaotic domains. As the energy goes beyond the escape energy,
 most chaotic orbits escape. In the first case we consider escapes to infinity,
 while in the second case we emphasize escapes to the central ``bumpy'' black
 hole. When the energy reaches its escape value a particular family of periodic
 orbits reaches an infinite period and then the family disappears (the orbit
 escapes). As this family approaches termination it undergoes an infinity of
 equal period and double period bifurcations at transitions from stability to
 instability and vice versa. The bifurcating families continue to exist beyond
 the escape energy. We study the forms of the phase space for various energies,
 and the statistics of the chaotic and escaping orbits. The proportion of these
 orbits increases abruptly as the energy goes beyond the escape energy.

\end{abstract}

\section{Introduction}\label{sec:intro}

The escapes from a dynamical system refer to a problem of basic interest
for dynamics: the problem of chaotic scattering. This problem has attracted the
attention of many authors in the previous decades (some typical
references are by
\cite{Churchill75,Petit86,Bleher88,Eckhardt88,Jung88,Henon89,Contop90,Contop92,
Ott93,Benet96,Benet98}).

In particular we found \citep{Contop04} the forms of the escape regions and the
escape rates in the simple dynamical system
\begin{equation} \label{func:Ham}
H=\frac{1}{2}(\dot{x}^2+\dot{y}^2+x^2+y^2)-xy^2=h
\end{equation}
for various values of the energy $h$ above the escape energy $h_{esc}=0.125$. We
found also regions of orbits that never escape, or escape after a very long time.
In particular the stable periodic orbits are surrounded by islands of stability
that never escape. This problem is typical of ``escapes to infinity". In fact,
similar results were found in general polynomial Hamiltonians representing two
perturbed harmonic oscillators.

In the present paper we explore the escapes in the classical system
(\ref{func:Ham}) and a very different problem of escapes, using a system from
the domain of General Relativity. In particular the system we use belongs to the
so-called Manko-Novikov (MN) metric family. \cite{ManNov92} found an exact
vacuum solution of Einstein's equations which describes a stationary,
axisymmetric, and asymptotically flat spacetime with arbitrary mass-multipole
moments. The MN metric subclass that we use can be considered as a perturbation
of the Kerr metric \citep{Kerr63} and it was introduced by \cite{Gair08}. The
Kerr metric represents a rotating black hole of mass $M$ and spin $S$. The MN
perturbation is expressed by a parameter $q$, which measures how much the MN
quadrupole moment $Q$ departs from the Kerr quadrupole moment $Q_{Kerr}=-S^2/M$
(i.e. $q=(Q_{Kerr}-Q)/M^3$). While the Kerr metric describes an integrable
system, the MN metric describes a non-integrable system that allows chaos. The
line element of the MN metric expressed in the Weyl-Papapetrou cylindrical
coordinates $(\rho, \varphi, z)$ is of the form
 \begin{equation} \label{func:MNmetric}
  ds^2=-f(dt-\omega d\varphi)^2 + f^{-1} [e^{2\gamma} (d\rho^2 + dz^2)
       +\rho^2 d\varphi^2],
 \end{equation}
where $f,~\omega,~\gamma$ are considered as functions of the prolate spheroidal
coordinates $v, w$, while the coordinates $\rho,z$ can be expressed as functions
of $v, w$ as well. Thus
\begin{equation}\label{func:trans}
 \rho=k \sqrt{(v^2-1)(1-w^2)},\quad z=k v w
\end{equation}
and
\begin{subequations}
\begin{eqnarray}
 f &=& e^{2 \psi}\frac{A}{B}, \label{ffunc} \\
 \omega &=& 2 k e^{-2 \psi}\frac{C}{A}-4 k \frac{\alpha}{1-\alpha^2}, \\
 e^{2 \gamma} &=& e^{2 \gamma^\prime}\frac{A}{(v^2-1)(1-\alpha^2)^2},
 \label{fexpgam} \\
 A &=& (v^2-1)(1+a~b)^2-(1-w^2)(b-a)^2,\label{fA} \\
 B &=& [(v+1)+(v-1)a~b]^2+[(1+w)a+(1-w)b]^2,\label{fB} \\
 C &=& (v^2-1)(1+a~b)[(b-a)-w(a+b)] \nonumber \\
   &&+ (1-w^2)(b-a)[(1+a~b)+v(1-a~b)], \\
 \psi &=& \beta \frac{P_2}{R^3}, \label{fC}\\
 \gamma^\prime &=& \ln{\sqrt{\frac{v^2-1}{v^2-w^2}}}+\frac{3\beta^2}{2 R^6}
 (P_3^2-P_2^2) \nonumber \\ &+& \beta \left(-2+\displaystyle{\sum_{\ell=0}^2}
 \frac{v-w+(-1)^{2-\ell}(v+w)}{R^{\ell+1}}P_\ell\right), \label{fgampr}\\
 a &=& -\alpha \exp {\left[-2\beta\left(-1+\displaystyle{\sum_{\ell=0}^2}
 \frac{(v-w)P_\ell}{R^{\ell+1}}\right)\right]}, \label{fa}\\
 b &=& \alpha \exp {\left[2\beta\left(1+\displaystyle{\sum_{\ell=0}^2}
 \frac{(-1)^{3-\ell}(v+w)P_\ell}{R^{\ell+1}}\right)\right]}, \label{fb}\\
 R      &=& \sqrt{v^2+w^2-1}, \label{fR}\\
 P_\ell &=& P_\ell (\frac{v~w}{R}). \label{fLegA}
\end{eqnarray}
\end{subequations}
Here $P_\ell(\zeta)$ is the Legendre polynomial of order $l$
\begin{equation} \label{fLeg}
 P_\ell(\zeta)=\frac{1}{2^\ell \ell!}
\left(\frac{d}{d\zeta}\right)^\ell(\zeta^2-1)^\ell,
\end{equation}
while the parameters $k,\alpha,\beta$ are related to the mass $M$, the spin $S$,
and the quadrupole deviation $q$ through the expressions
\begin{equation}
\begin{array}{r}
\alpha=\frac{-1+\sqrt{1-\chi^2}}{\chi},
\end{array}
\begin{array}{c}
k=M\frac{1-\alpha^2}{1+\alpha^2},
\end{array}
\begin{array}{l}
\beta=q \left( \frac{1+\alpha^2}{1-\alpha^2} \right)^3.
\end{array}
\label{freepar}
\end{equation}
while $\chi$ is the dimensionless spin parameter $\chi=S/M^2$. These formulae
give the Kerr metric when $q=0$.

Contrary to the system (\ref{func:Ham}), the escapes in the MN system refer not
only to infinity, but also to orbits that fall into the central compact object
called ``bumpy black hole'' by \cite{Gair08}. The MN central compact object is
not really a black hole, because its horizon is broken along the equator and
regions of closed timelike curves exist outside the horizon \citep{Gair08}. The
MN problem is very different from the simple Hamiltonian (\ref{func:Ham});
however, the two problems have several common properties concerning the role of
the periodic orbits and their relations to escapes.

In the present paper we study in detail the periodic orbits in the systems
(\ref{func:Ham}) and (\ref{func:MNmetric}) and their connection with the escape
phenomena. We found that the periodic orbits in general do not lead to escapes,
except in very special cases. Most of the families of periodic orbits generate
an infinity of unstable, but non-escaping, periodic orbits. On the other hand,
the asymptotic curves of the various unstable periodic orbits play an important
role in generating chaos and most chaotic domains lead to escapes when the
energy goes beyond the escape energy.

The paper is organized as follows. In section \ref{sec:PerOrbHam} we calculate
the periodic orbits of the system (\ref{func:Ham}). Then we find the asymptotic
curves of the unstable periodic orbits and the corresponding chaos. In section
\ref{sec:EscHam} we find the escapes from the system (\ref{func:Ham}) and their
statistics. In section \ref{sec:PerOrbMN} we find the periodic orbits in the
system (\ref{func:MNmetric}) and the chaotic domains. In section \ref{sec:EscMN}
we find the corresponding escapes and their statistics. Finally, in section
\ref{sec:concl} we compare the two systems and draw our main conclusions.

\section{Periodic Orbits in the system (\ref{func:Ham})} \label{sec:PerOrbHam}

\begin{figure}[htp]
 \centerline{\includegraphics[width=0.51\textwidth] {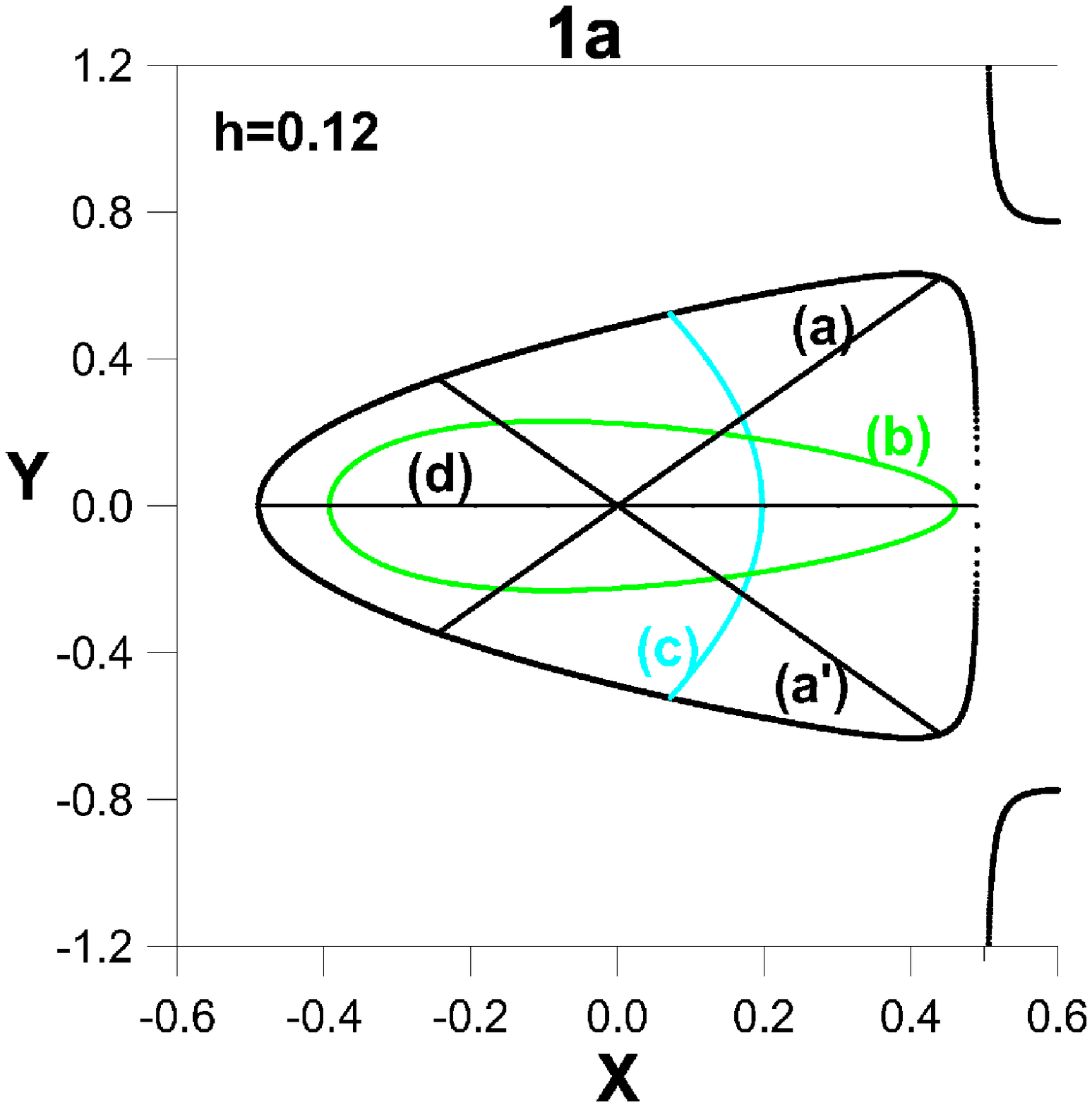}
 \includegraphics[width=0.49\textwidth] {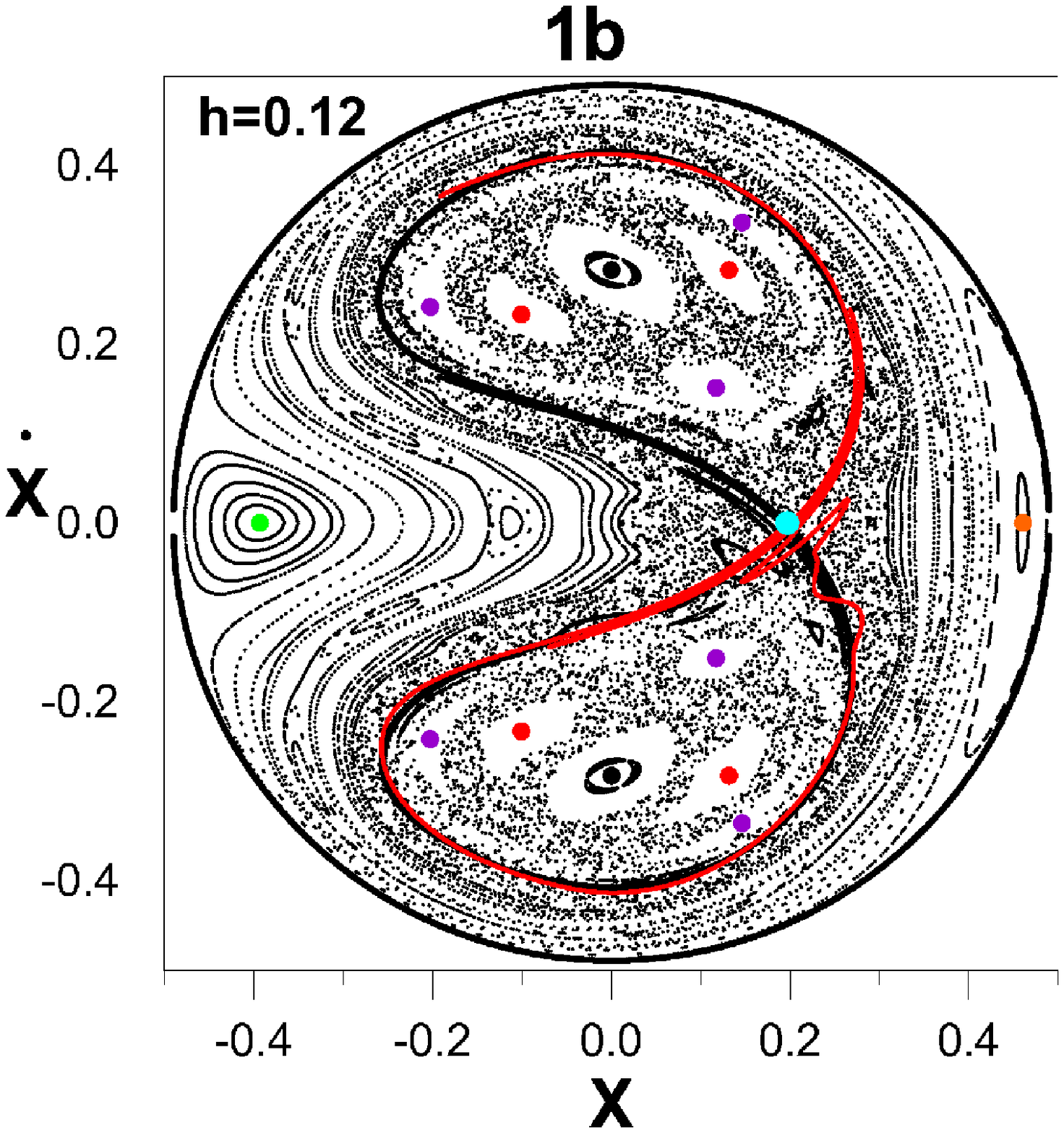}}
 \caption{
  (\ref{fig:01}a) Simple periodic orbits (of period 1) for energy $h=0.12$.
  (\ref{fig:01}b) A surface of section $(x, \dot{x})$ for $h=0.12$~ containing
  ordered and chaotic orbits. On the $\dot{x}=0$ axis there are two stable
  periodic orbits ($b$) and an unstable orbit ($c$) (at the intersection of the
  red and black curves), together with one stable (red) and one unstable (thick
  black) asymptotic manifolds. There are also two stable orbits (a) on the $x=0$
  axis and triple and double period stable orbits that have bifurcated from the
  orbits ($a$) and ($a'$) on the $x$=0 axis.
}
\label{fig:01}
\end{figure}

For small values of the energy the system (\ref{func:Ham}) has 4 periodic orbits
of period 1. There are two straight line periodic orbits
 \begin{equation}\label{func:StrLinPO}
  y=\pm\sqrt{2}~x
 \end{equation}
that are stable (orbits ($a$) and ($a'$) in Fig. \ref{fig:01}a). There is also a
stable orbit ($b$), like an ellipse, described clockwise or counterclockwise, a
simple unstable orbit crossing the $x-axis$ perpendicularly (orbit ($c$)), and
the orbit $y=0$ (orbit (d)). In Fig. \ref{fig:01}b we see also $2$ stable triple
orbits $3:1$ and $2$ stable double orbits $2:1$ that have bifurcated from the
orbits ($a$) and ($a'$).

The orbits $y=\pm\sqrt{2}x$ undergo an infinity of transitions to instability
and stability  as \emph{h} tends to $h_{esc}$ while the period of these orbits
tends to infinity \citep{Contop80}. The periodic orbits $a$ and $a'$ terminate
at $h=h_{esc}$ when their period is infinite, in accordance with the
Str\"{o}mgren termination principle \citep{Szebehely67}. For $h>h_{esc}$ the
Curve of Zero Velocity (CZV) $(x^2+y^2-2~x~y^2=2h)$ opens and the orbits $a$ and
$a'$ escape to infinity.

The families ($a$) and ($a'$) at every transition to stability or instability
generate by bifurcation new families of equal or double period. Families of
higher order are also bifurcated from the families ($a$) and ($a'$).

On a surface of section $(x,\dot{x})$ with $(y=0 ~and~ \dot{y}>0)$ the stable
orbits are surrounded by islands of stability, while the unstable orbit is
surrounded by a  set of chaotic orbits (Fig. \ref{fig:01}b). In particular the
orbits ($a$) and ($a'$) appear on the axis $x=0$ of Fig. \ref{fig:01}b. Near the
unstable orbit ($c$) there is a chaotic region (Fig. \ref{fig:01}b).

As $h$ increases the chaotic region around the unstable orbit (c) increases,
while various sets of higher order periodic orbits bifurcate from the straight
line orbits (($a$), ($a'$)) (Fig. \ref{fig:01}b).

\begin{figure}[htp]
 \centerline{\includegraphics[width=0.7\textwidth] {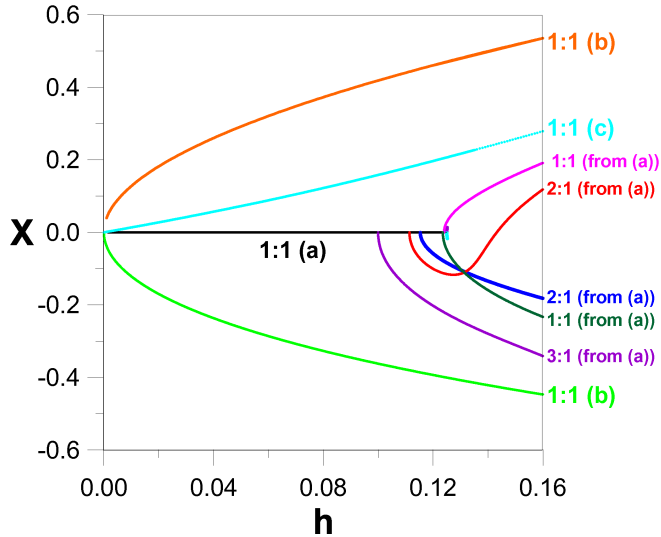}}
 \caption{Characteristics of some of the most simple families of periodic orbits.
  The family (a) and its bifurcations have $\dot{x}\neq 0$
}
\label{fig:02}
\end{figure}

\begin{figure}[htp]
 \centerline{\includegraphics[width=0.495\textwidth] {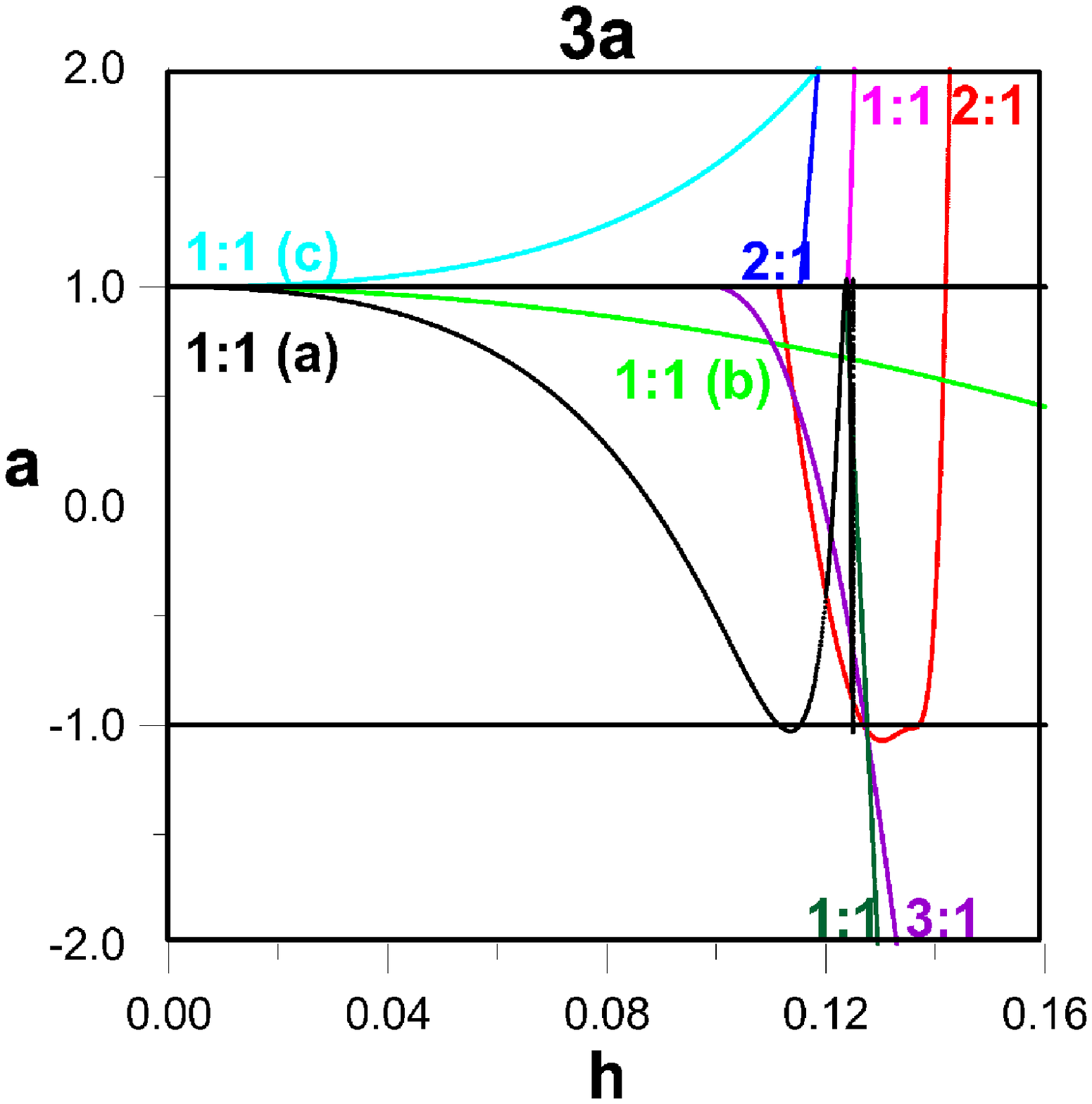}
 \includegraphics[width=0.505\textwidth] {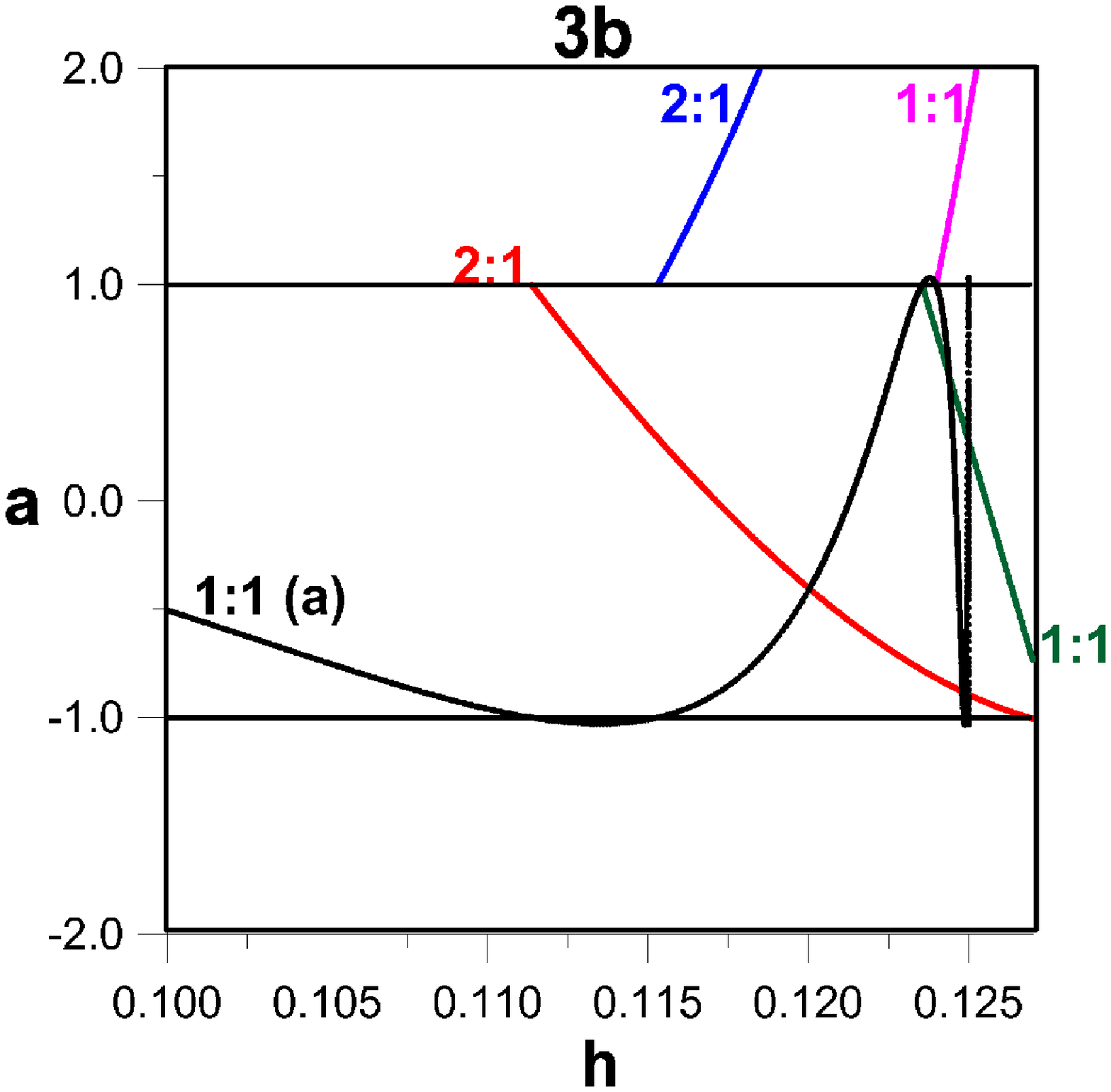}}
 \caption{
 (\ref{fig:03}a) The H\'{e}non stability parameter \emph{a}~of the families of
 Fig. \ref{fig:02} as a function of the energy \emph{h}; (\ref{fig:03}b) Details
 of Fig. \ref{fig:03}a.
}
\label{fig:03}
\end{figure}

The characteristics of the various families of periodic orbits are shown in Fig.
\ref{fig:02}. These characteristics give the value of \emph{x} when $y=0$ and
$\dot{y}>0$. We mark in particular the orbits $3:1$, $2:1$ and $1:1$ bifurcating
from the orbits ($a$) and ($a'$) as the energy $h$ increases. The stability of
these families is given in Fig. \ref{fig:03}. The orbits are stable if their
H\'{e}non parameter $a$ is between $-1$ and $+1$.

The family $y=\sqrt{2}~x$ (a) is stable for $0<h<0.111$, then unstable in the
interval $0.111<h<0.116$, stable again for $0.116<h<0.1236$, and has infinite
more transitions to instability and stability until $h=h_{esc}=0.125$. At the
first transition to instability $(h=0.111)$ the value of \emph{a} goes below
$a=-1$ and one stable family of double period bifurcates there. Its stability is
represented by the red curve in Fig. \ref{fig:03}, starting at $a=1$.

At the first transition from instability to stability along the family (a)
$(h=0.116)$ \emph{a} goes above $a=-1$ and there is a bifurcation of an unstable
family $2:1$ (blue curve above $a=1$).

At the second transition to instability the value of \emph{a} goes above $a=1$
$(h=0.1236)$ and two  different stable families of equal period bifurcate there
(black 1:1). The orbits of these families deviate from the straight line
$y=\sqrt{2}~x~$. At the second transition to stability there are bifurcations of
two unstable families (magenta 1:1), and so on.

Symmetric bifurcations are formed also around the family ($a'$). There are also
higher order bifurcations from the families ($a$) and ($a'$), e.g. the family
$3:1$ (Figs. \ref{fig:02} and \ref{fig:03}a).

\begin{figure}[htp]
 \centerline{\includegraphics[width=0.51\textwidth] {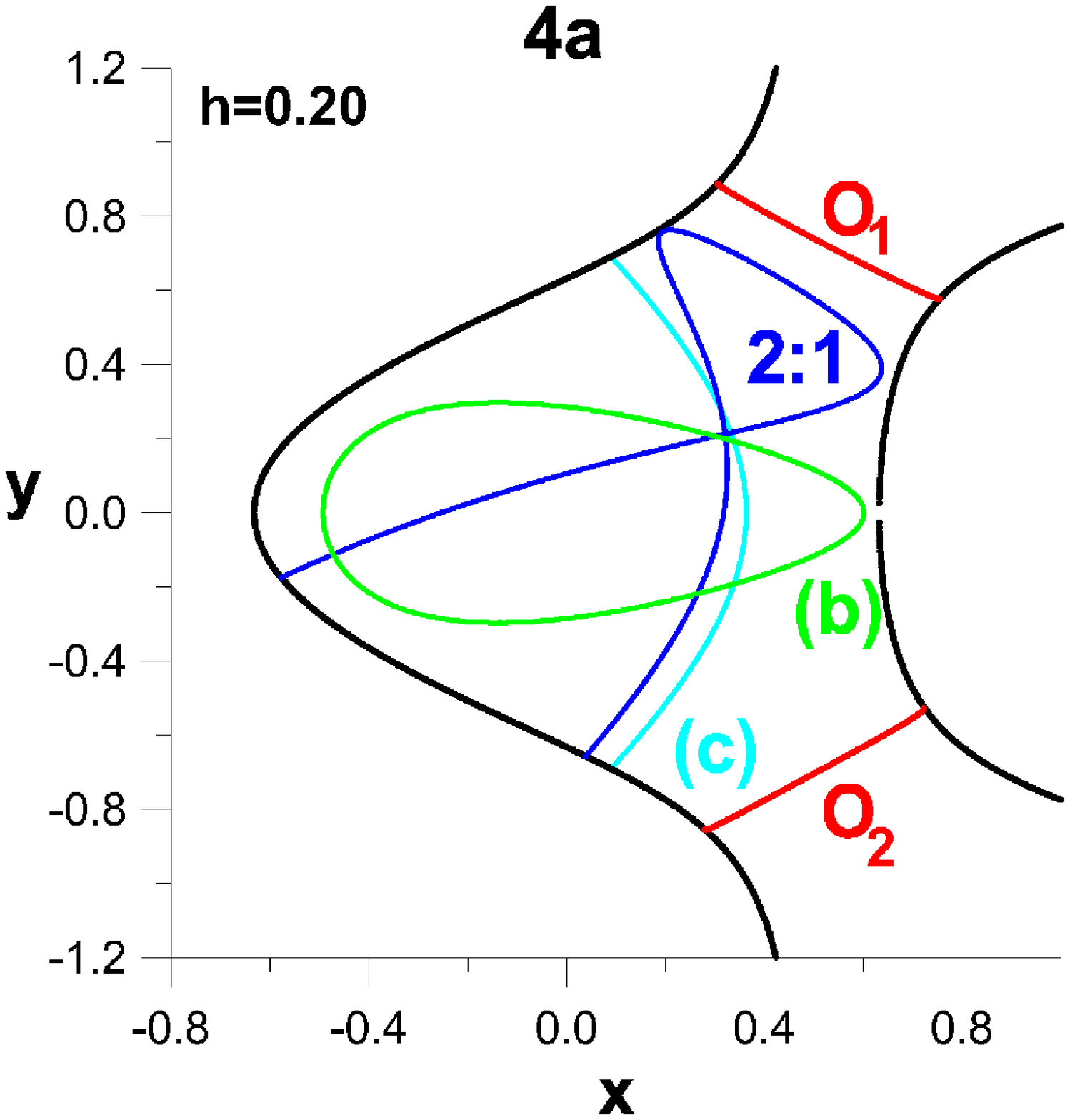}
 \includegraphics[width=0.49\textwidth] {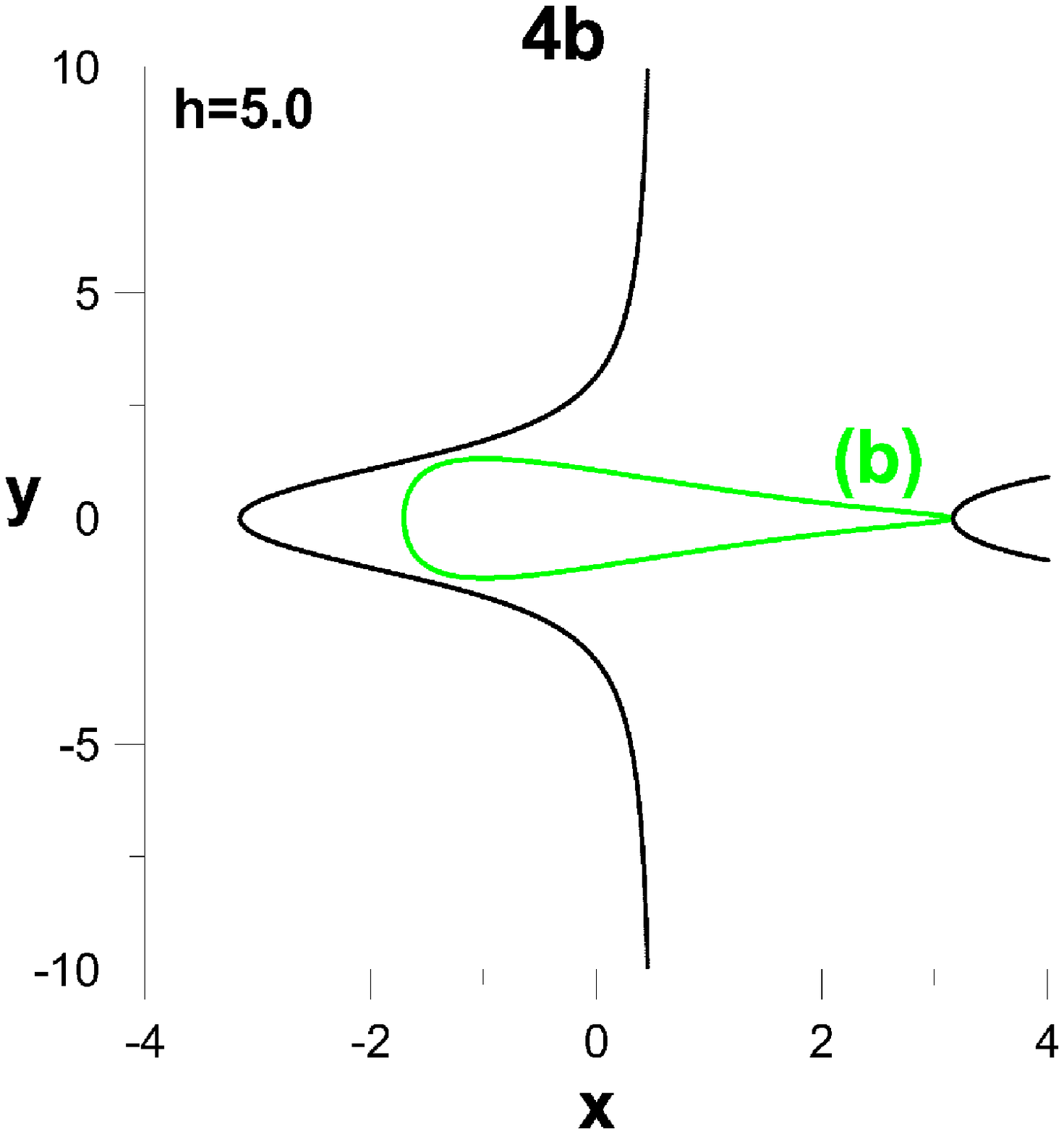}}
 \caption{
  (\ref{fig:04}a) Some periodic orbits for $h=0.2$. Four periodic orbits of
  period 1, ($b$), ($c$), and the Lyapunov orbits $O_1$, $O_2$, and a periodic
  orbit of period 2, belonging to the family bifurcating from the family ($a$).
  (\ref{fig:04}b) An (unstable) periodic orbit ($b$) for $h=5.0$.
}
\label{fig:04}
\end{figure}

The  families bifurcating from the orbit (a) consist of orbits that reach the
CZV at two points. These families do not terminate for $h$ larger than
$\emph{h}=0.125$, but extend all the way to $h=\infty$. An example of an orbit
of such a family $(2:1)$ is shown in Fig. \ref{fig:04}a for $h=0.2$. This family
is bifurcated from the family (a) for $h=0.111$ and it is stable in the interval
$0.111<h<0.1252$. At $\emph{h}=0.1252$ the family $2:1$ becomes unstable and
generates a family with double period (i.e. 4 times the original period). For
$h=0.137$ the family $2:1$ crosses again the. $a=-1$ axis (Fig. \ref{fig:03}a)
and becomes again stable generating an unstable double period family (i.e. 4
times the original period). Finally, at $\emph{h}=0.142$~the  family $2:1$
crosses the $a=1$ axis and becomes unstable for larger \emph{h}. At this
crossing point it generates an equal period stable family (2 times the original
period). This also becomes unstable for a slightly larger \emph{h}, followed by
a cascade of period doubling bifurcations, that generate an infinity of unstable
families, that exist for arbitrarily large \emph{h}.

When $h>h_{esc}=0.125$ the CZVs open above and below the axis $(y=0)$ (Fig.
\ref{fig:04}a,b) and several orbits escape to infinity. However, the periodic
orbits of Figs. \ref{fig:04}a,b do not escape. E.g. the orbit $2:1$ in Fig.
\ref{fig:04}a that bifurcated from the orbit (a) at $h=0.111$ does not escape,
although the orbit (a) has escaped. In Fig. \ref{fig:04}b we show a non-escaping
orbit (b) for $h=5.0$ when the openings are very large.

\begin{figure}[htp]
 \centerline{\includegraphics[width=0.7\textwidth] {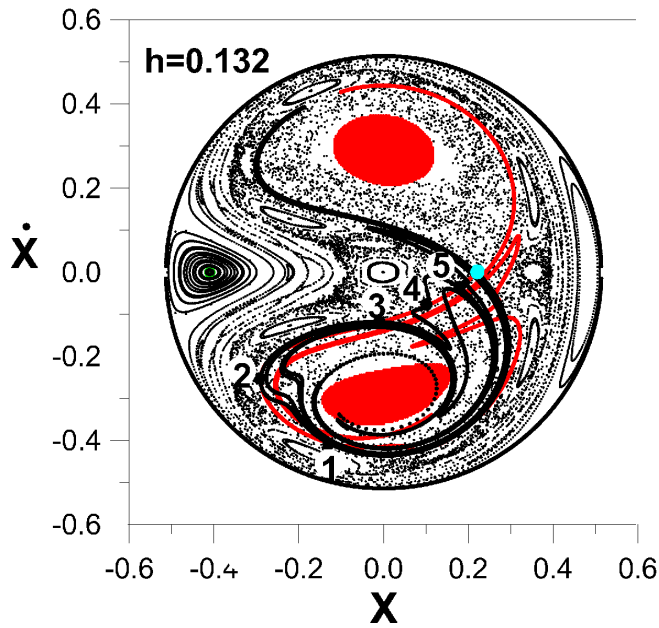}}
 \caption{ Surface of section for $h=0.132$. The orbits in the red regions
  escape directly to infinity without any intersection with the axis $y=0$. Blue
  dot corresponds to the orbit (c).
}
\label{fig:05}
\end{figure}

At every opening of the CZV there is an unstable orbit (called Lyapunov orbit)
crossing it (Fig. \ref{fig:04}a). Any orbit, crossing a Lyapunov orbit outwards,
escapes to infinity \citep{Churchill75,Contop04}. The regions of direct escape
are shown in red in Fig. \ref{fig:05} for $h=0.132$. Orbits starting in these
regions escape directly, without any intersections with the axis $y=0$. The
boundaries of the direct escape regions are defined by the asymptotic curves of
the Lyapunov orbits. These asymptotic curves make infinite rotations around the
escape regions \citep{Contop90}. In that paper we described in detail the
asymptotic curves from the Lyapunov orbits. We found that these asymptotic
curves include an infinity of branches that start and end with infinite
rotations around the escape regions. The stable asymptotic curves of any
other unstable periodic orbit leading to escapes approach the same limiting
asymptotic curves, i.e. the boundaries of the red regions as in Fig.
\ref{fig:05}. For relatively small values of the energy $h$, however, most of
the chaotic orbits escape after one or more intersections with the axis $y=0$
(Fig. \ref{fig:05}).

The asymptotic curves of the unstable orbit ($c$) surround the red escape
regions. Examples of such  asymptotic curves are shown in Figs. \ref{fig:01}b
and \ref{fig:05}. The unstable asymptotic curves start in opposite directions
and the stable asymptotic curves start also in opposite directions. In Figs.
\ref{fig:01}b and \ref{fig:05} we give in black the unstable asymptotic curve
that starts towards the right downwards and surrounds clockwise the lower red
region. As it comes again close to the point ($c$) it makes several oscillations,
up and down close to the point ($c$).

The stable asymptotic curve (red) starts to the left and downwards, surrounding
counterclockwise the red escape region and intersects the unstable asymptotic
curve at several homoclinic points. The oscillations of the stable and unstable
asymptotic curves are much larger in Fig. \ref{fig:05} $(h=0.132)$ than in Fig.
\ref{fig:01}b $(h=0.12)$. But the main difference between Figs. \ref{fig:01}b
and \ref{fig:05} is that in the first case there are no escapes, while in the
second case most of the chaotic orbits escape to infinity. The orbits that
escape directly without any intersection with the axis $y=0$, are marked in red.
The stable asymptotic curve (red curve) of the unstable periodic orbit (c) (blue
dot in Fig. \ref{fig:05}) makes an infinity of rotations around the lower red
region. This asymptotic curve never intersects the red regions. This happens
because orbits starting on the stable asymptotic curve have further
intersections along the same asymptotic curve, until they reach asymptotically
the periodic orbit. Therefore they cannot escape immediately, hence they cannot
be found inside a red region.

On the other hand, the unstable asymptotic curve (black curve) intersects the
red region after a number of oscillations below and above the stable asymptotic
curve (Fig. \ref{fig:05}). The stable and unstable asymptotic curves intersect
at an infinity of homoclinic points. A few points $(1-5)$ are marked in Fig.
\ref{fig:05}. The corresponding homoclinic orbits approach asymptotically the
periodic  orbit (c) as $t\rightarrow\infty$ and $t\rightarrow -\infty$. The
unstable asymptotic curve (black) makes outer and inner loops above and below
the stable asymptotic curve. The areas of these lobes are equal. The homoclinic
intersections approach closer and closer the periodic orbit (c), their distances
decreasing proportionally to $1/\lambda$, where $\lambda$ is the larger
(positive) eigenvalue of the orbit (c). Thus the lengths of the lobes become
longer proportionally to $\lambda$. The value of $\lambda$ increases as the
energy increases.

The inner lower lobes make some rotations, back and forth, around the lower red
escape region. The inner loop starting downwards from the homoclinic point 5
makes $1/2$  rotation around the red region and returns closer than the point 5
to the periodic orbit (c). The next inner loop makes about $1.5$ rotations
around the red region and the subsequent loop intersects the red region (Fig.
\ref{fig:05}).

Then the orbits passing through points  of the black curve inside the red region,
escape to infinity without any further intersections with the surface of section.
As a point along the black curve approaches the red region its next image on the
surface of section makes an infinity of rotations around the upper red region
of Fig. \ref{fig:05} approaching asymptotically the boundary of the red region.
This phenomenon is described in detail in the paper of \cite{Contop04}.

For values of $h$ up to about $0.32$ there are some stable families, generated
from the stable family ($b$), of orbits starting on the $y=0$ axis. This family
consists of almost elliptical orbits that are described either clockwise or
counterclockwise (Figs. \ref{fig:01}a,b). The family ($b$) is stable up to about
$h=0.304$ and then it becomes unstable at a period doubling bifurcation ($a=-1$).
The  double period families bifurcating from ($b$) start at $a=+1$ and become
unstable at another period doubling bifurcation (4 times the original period)
for about $h=0.32$. Thus the interval between the first and the second period
doubling bifurcations is $\Delta h =0.016$. This is followed by a cascade of
period doubling bifurcations that produce an infinity of unstable families,
existing for arbitrarily large $h$. It is known that the intervals between
successive period doubling bifurcations in conservative systems decrease by a
universal (asymptotic) factor $\delta=\frac{1}{8.2})$ \citep{Eckmann81}.
Therefore, the total interval $\Delta h$ that contains higher order bifurcation
of the family (b) is approximately $\Delta=\frac{0.016}{1-\frac{1}{8.2}}=0.018$,
and beyond $h=0.304+0.018=0.322$ there is no stable periodic orbit generated by
bifurcations from the family (b).

\begin{figure}[htp]
\centerline{\includegraphics[width=0.6\textwidth] {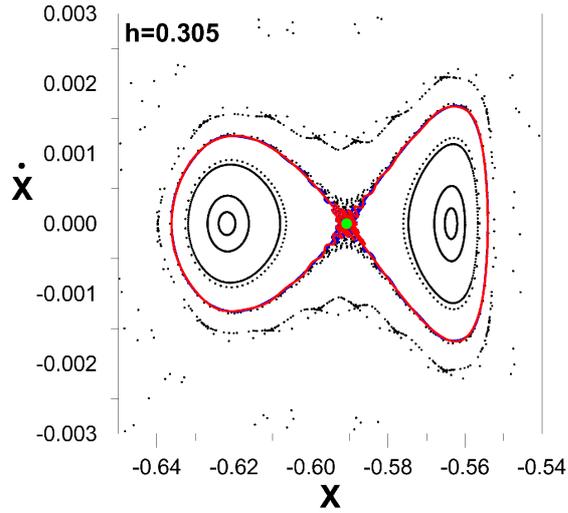}}
 \caption{ The periodic orbit ($b$) on the surface of section for $h=0.305$,
  when the periodic orbit has just become unstable, and two stable orbits that
  bifurcated from it. The asymptotic curves from the orbit ($b$) surround both
  islands of stability, but they do not extend to larger distances.}
\label{fig:06}
\end{figure}

\begin{figure}[htp]
 \centerline{\includegraphics[width=0.51\textwidth] {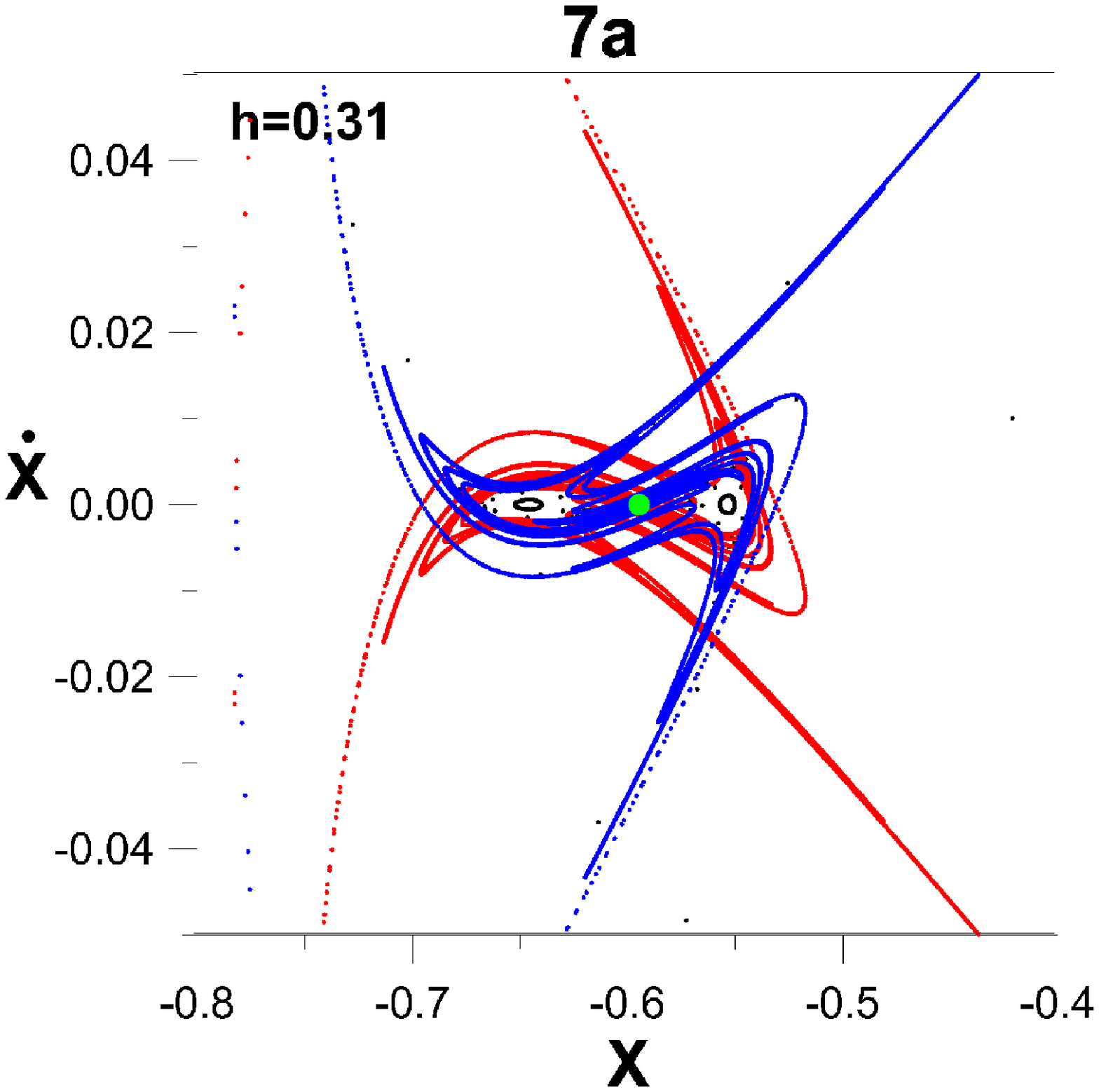}
 \includegraphics[width=0.49\textwidth] {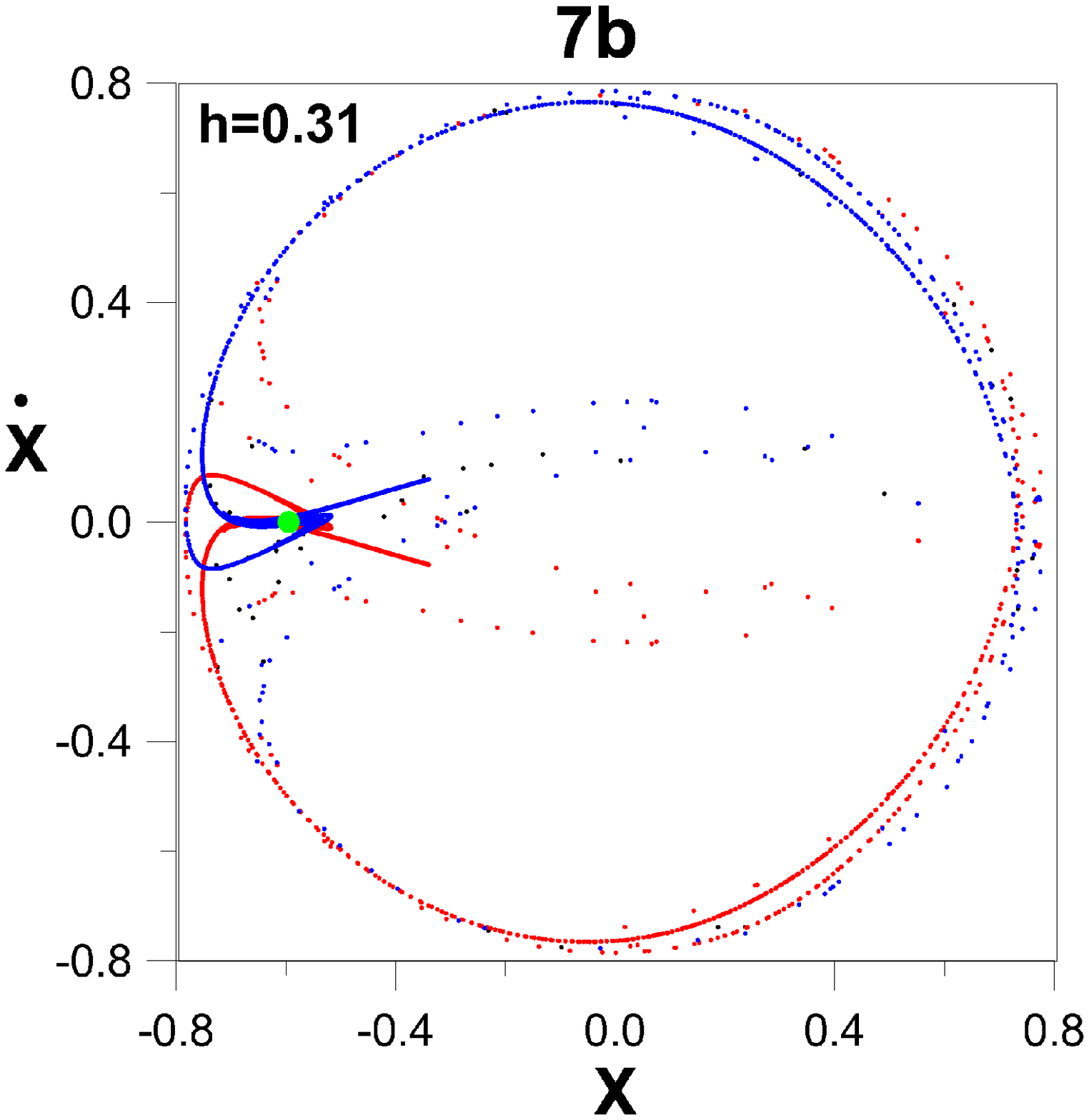}}
 \caption{ The asymptotic curves (blue=unstable and red=stable) of the orbit
 (b) for $h=0.31$ extend well beyond the areas around the two islands
 (Fig. \ref{fig:07}a) and go all around the available area on the surface of
  section (Fig. \ref{fig:07}b).
}
\label{fig:07}
\end{figure}

When the orbit ($b$) has just become unstable there are two islands of stability
around it. For $h=0.305$ (Fig. \ref{fig:06}) the asymptotic curves of the orbit
($b$) surround the two islands, but they do not extend to large distances. In
fact, for this value of \emph{h} there are invariant curves surrounding the
orbit ($b$) and the islands around it. However, for still larger \emph{h} the
asymptotic curves of ($b$) extend very far (Fig. \ref{fig:07}a) surrounding the
whole available area on the surface of section (Fig. \ref{fig:07}b). The
invariant curves surrounding the point (b) for $h=0.305$, have been destroyed
for $h=0.31$ and the asymptotic curves come close to the limiting curve
$\dot{x}^2+x^2=2h$.

The same pattern of transition to instability is followed by most stable
periodic orbits. Namely most stable orbits become unstable at a period doubling
bifurcation followed by a cascade of infinite period doublings that lead to an
infinity of unstable periodic orbits. An exception are the families ($a$) ($a'$),
which have a non-universal bifurcation ratio $\delta =9.22$ and lead to escapes
\citep{Contop80}. A theoretical explanation of this particular bifurcation ratio
was provided by \cite{Heggie83}. A review of previous theoretical and numerical
work on bifurcations is provided by \cite{Contop02}.

\section{Escapes from the system (\ref{func:Ham})} \label{sec:EscHam}

As the energy $h$ increases the proportion of orbits, that escape directly
through the openings of the the CZV increases. In Fig. \ref{fig:05} $(h=0.132)$
the upper red region represents orbits that escape through the upper opening of
the CZV, and the lower red region represents orbits that escape through the
lower opening of the CZV, without ever intersecting the axis $y=0$. However,
most of the chaotic orbits, represented by scattered dots in Fig. \ref{fig:05},
escape from the system after a small or large number of intersections with the
$y=0$ axis.

\begin{figure}[htp]
 \centerline{\includegraphics[width=0.7\textwidth] {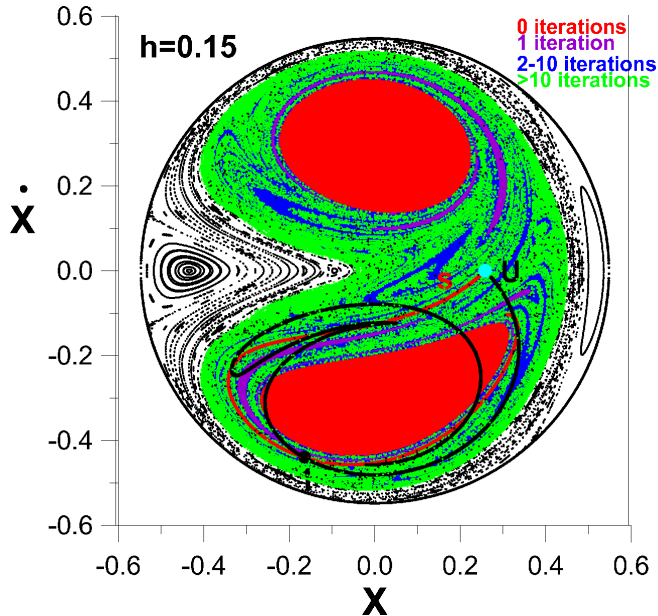}}
 \caption{ Escape regions for $h=0.15$.
}
\label{fig:08}
\end{figure}

As $h$ increases to $h=0.15$ the red regions increase in relative size
(proportion of the total area). In Fig. \ref{fig:08} are shown the direct escape
regions for $h=0.15$ and the asymptotic curves (black, unstable and red, stable),
from the periodic orbit ($c$), in the lower part of the figure. The regions in
Fig. \ref{fig:08} are colored according to the number of intersections an orbit
has with the axis $y=0$ before the orbit escapes. The stable asymptotic curve
(red) again does not intersect the red region, but the (black) unstable
asymptotic curve intersects the red region on the left after half a rotation
around it. Then it intersects it again (after an inner and an outer loop, below
and above the stable asymptotic curve) on the right side of the red region and
so on. The orbits starting at points of the black curve inside the red region
escape again without any further intersections with the axis $y=0$.

\begin{figure}[htp]
 \centerline{\includegraphics[width=0.495\textwidth] {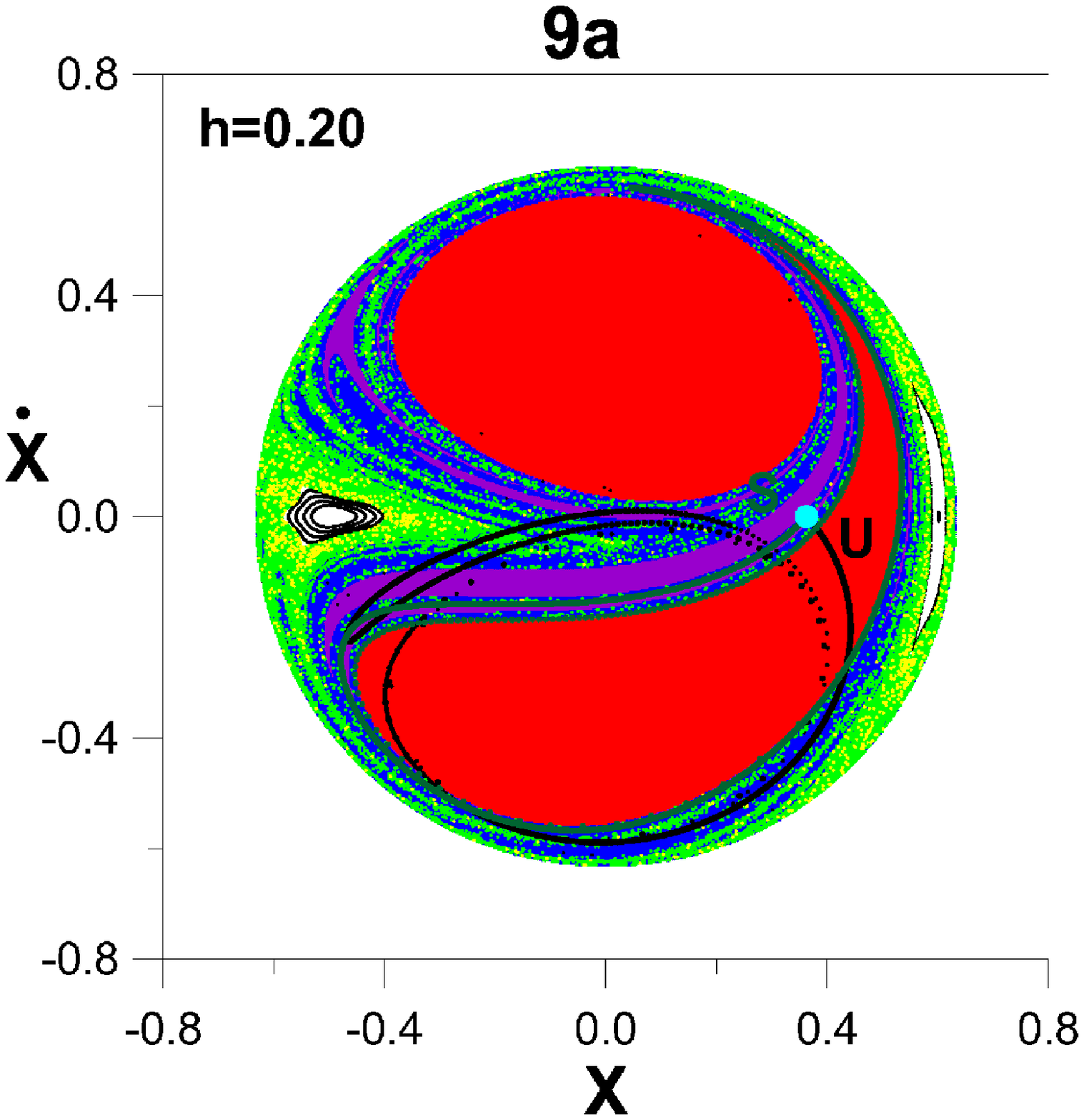}
 \includegraphics[width=0.505\textwidth] {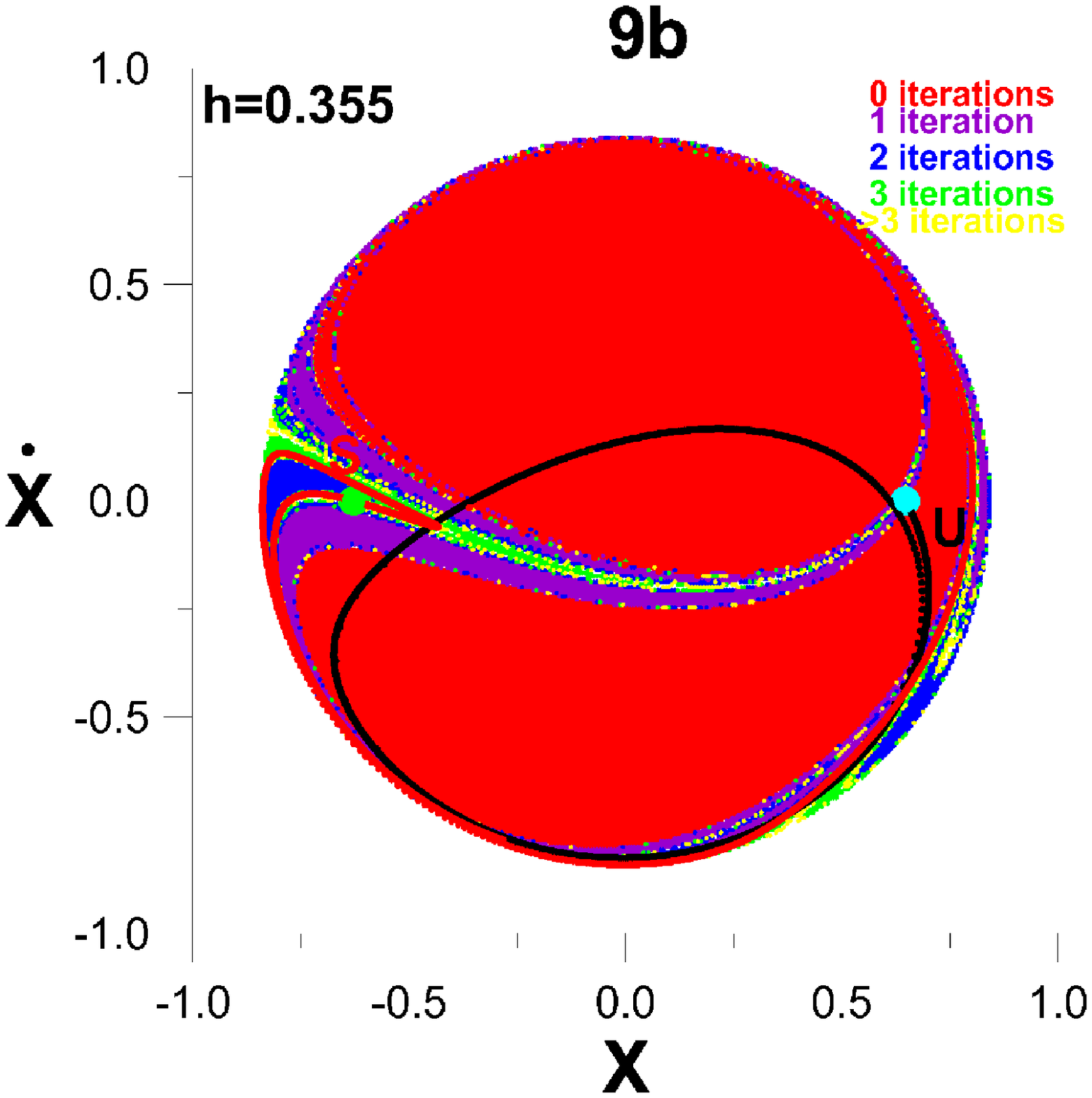}}
 \caption{
 (\ref{fig:09}a) Escape regions for $h=0.25$ either directly (red), or after a
 number of intersections of the $y=0$ axis upwards $(\dot{y}>0)$, as indicated
 in the color scale. The white region represents an island of stability. We mark
 also an unstable asymptotic curve ($U$) from the periodic orbit ($c$).
 (\ref{fig:09}b) The same, as in Fig. \ref{fig:09}a, for $h=0.355$, together
 with an initial arc of the stable asymptotic curve of the orbit ($b$) on the
 left (red).
}
\label{fig:09}
\end{figure}

For larger \emph{h} ($h=0.20$) (Fig. \ref{fig:09}a) and $h=0.355$ (Fig.
\ref{fig:09}b) the two regions of direct escapes increase in proportion to the
total available space.

In Fig. \ref{fig:09}b we have marked one unstable asymptotic curve of the
unstable periodic orbit (c) (black), and one stable  asymptotic curve of the
unstable periodic orbit (b) on the left (red). The two curves are almost tangent
at a point. We have checked that for $h=0.354$ the two asymptotic curves do not
intersect  and for $h=0.356$ they clearly intersect close to this point. Thus,
for a value of $h$ a little smaller than $h=0.355$ the two curves are tangent.
Then according to the Newhouse theorem \citeyearpar{Newhouse77,Newhouse83} close
to the tangency point there is a stable periodic orbit, generated at a tangent
bifurcation.

In Figs. \ref{fig:09}a,b we see that the periodic orbits ($b$) and ($c$) are in
a region where escapes occur after several intersections of the axis $y=0$ by
the orbits. In fact, the periodic orbits never escape, and orbits close to them
escape after a time that increases to infinity as the initial conditions
approach the periodic orbits. In particular the stable asymptotic curves do not
intersect the red regions of direct escapes. In fact, all the orbits starting on
the stable asymptotic curves never escape.

On the other hand, the unstable asymptotic curve $U$ of ($c$) in Figs.
\ref{fig:09}a,b enters the red region at a rather small distance from the orbit
($c$), and the orbits starting inside the red region escape directly. However,
this unstable asymptotic curve comes out of the red region several times, hence
the corresponding orbits escape in general after some  intersections with the
axis $y$=0.

The fact that there are chaotic orbits that escape after long times is related
to the phenomenon of stickiness along the unstable asymptotic curves
\citep{ContopHar08,ContopHar10}.

As $h$ increases further new tangencies between the unstable and the stable
asymptotic curves of the unstable orbits are formed, thus new stable periodic
orbits appear. However, the islands around these periodic orbits are extremely
small, since almost $100\%$ of the orbits escape for large $h$.

This situation is similar to the case where no escapes exist at all, but
tangencies between the asymptotic curves generate new stable periodic orbits
\citep{Contop94}.

\begin{figure}[htp]
 \centerline{\includegraphics[width=0.495\textwidth] {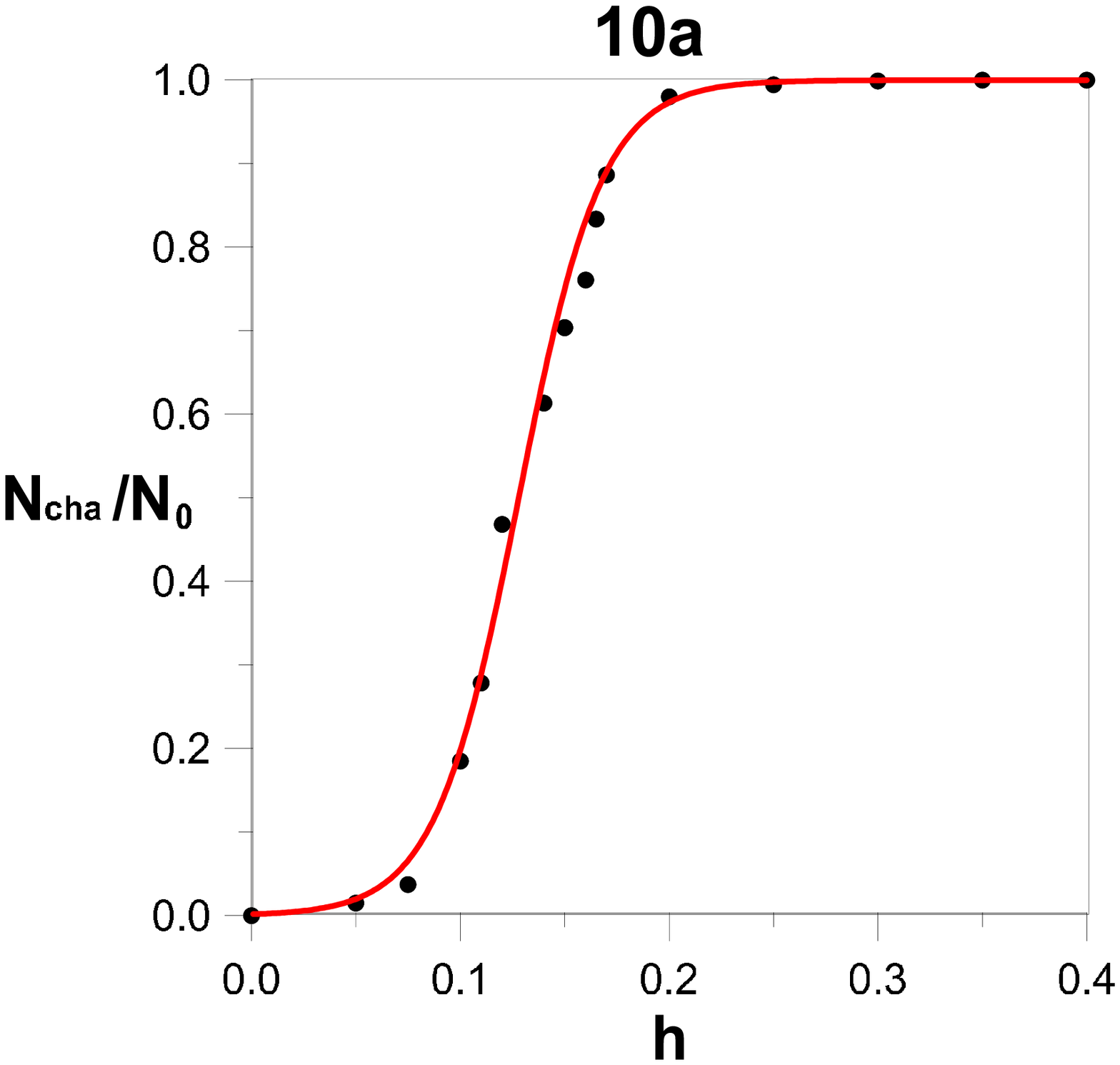}
 \includegraphics[width=0.505\textwidth] {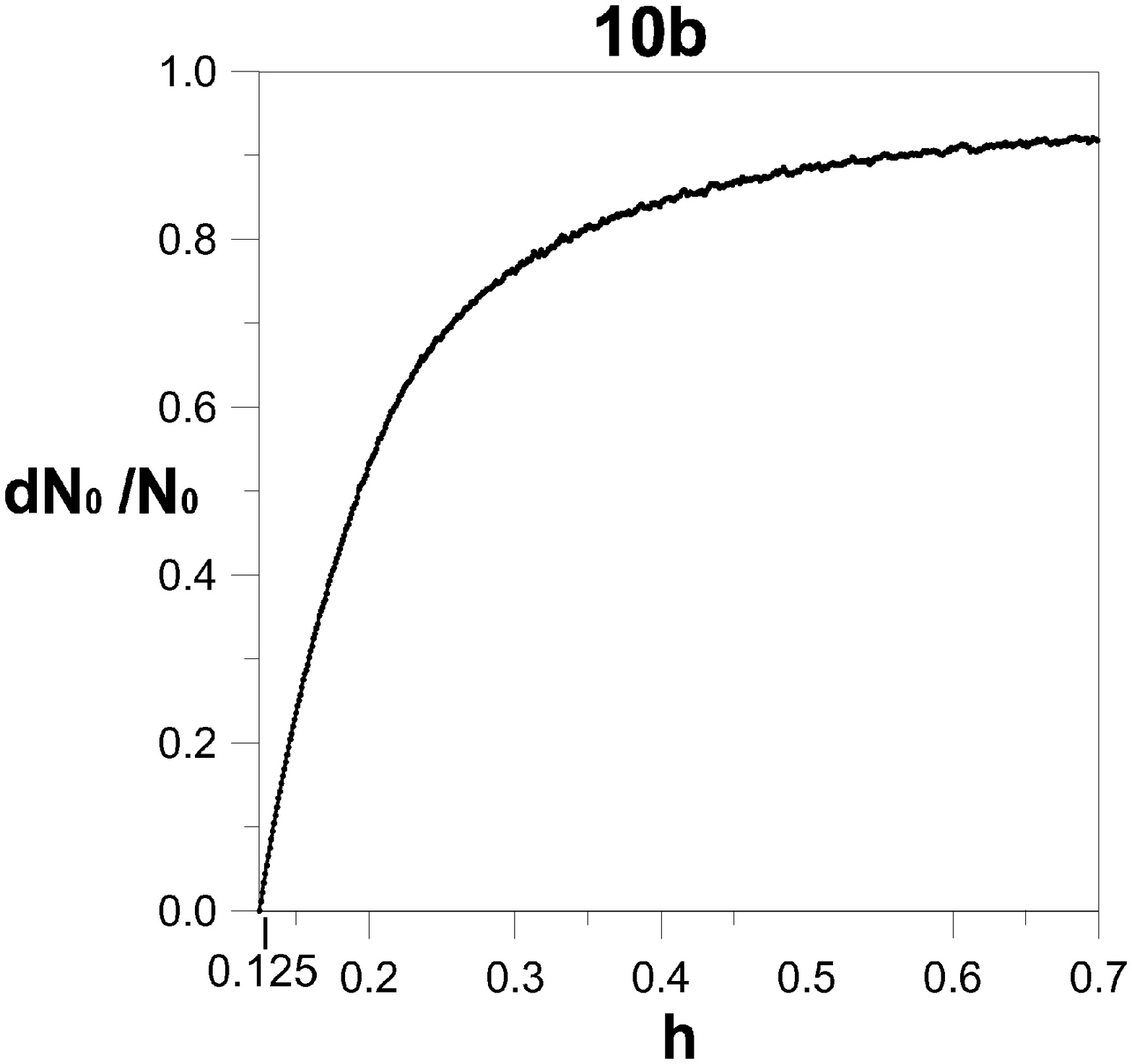}}
 \caption{
 (\ref{fig:10}a) The proportion of chaotic (including escaping) orbits
 $N_{cha}/N_0$ as a function of the energy $h$.
 (\ref{fig:10}b) The proportion of the directly escaping orbits $dN_0/N_0$
 (without any intersection with the axis $y=0$) as a function of the energy h.}
\label{fig:10}
\end{figure}

\begin{figure}[htp]
 \centerline{\includegraphics[width=0.495\textwidth] {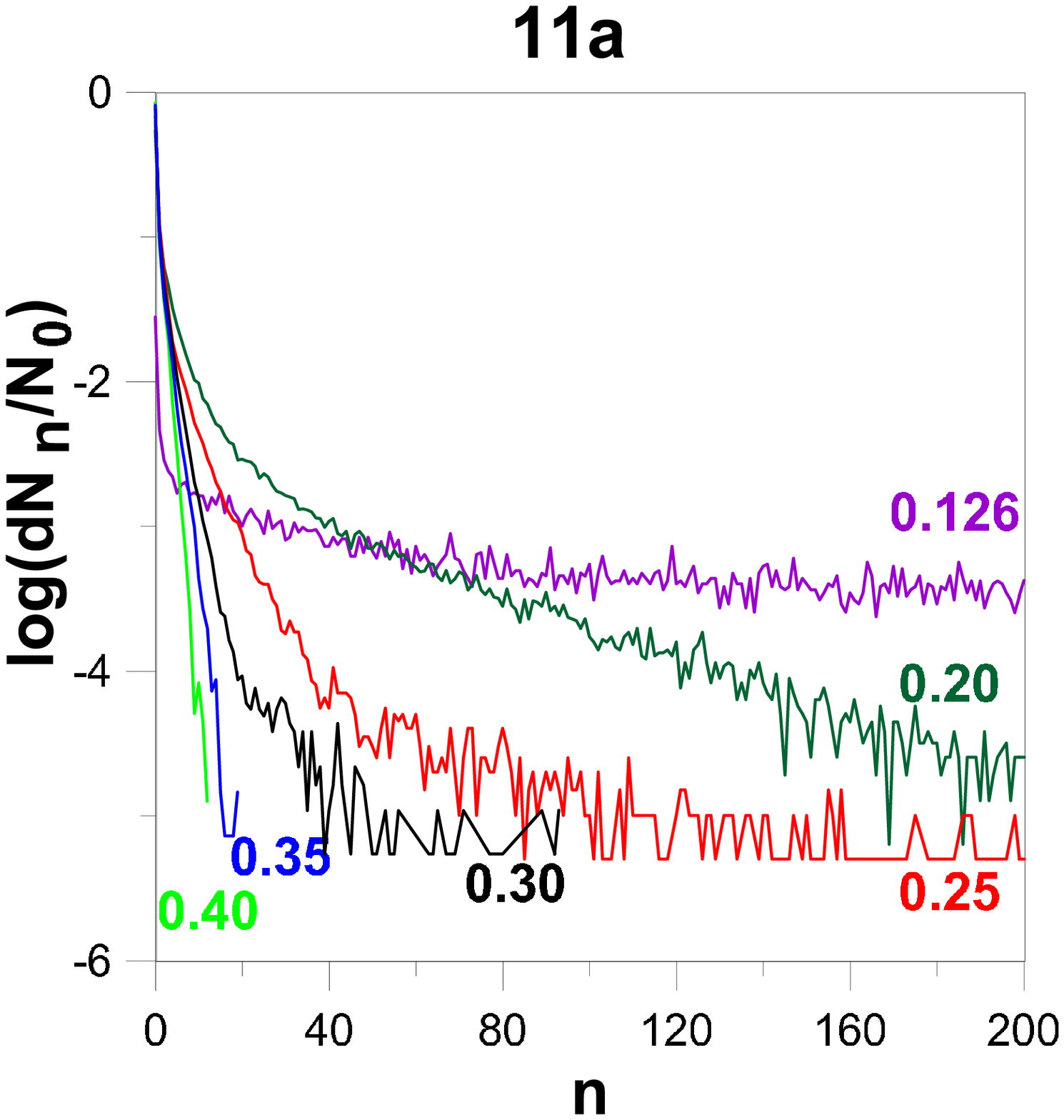}
 \includegraphics[width=0.505\textwidth] {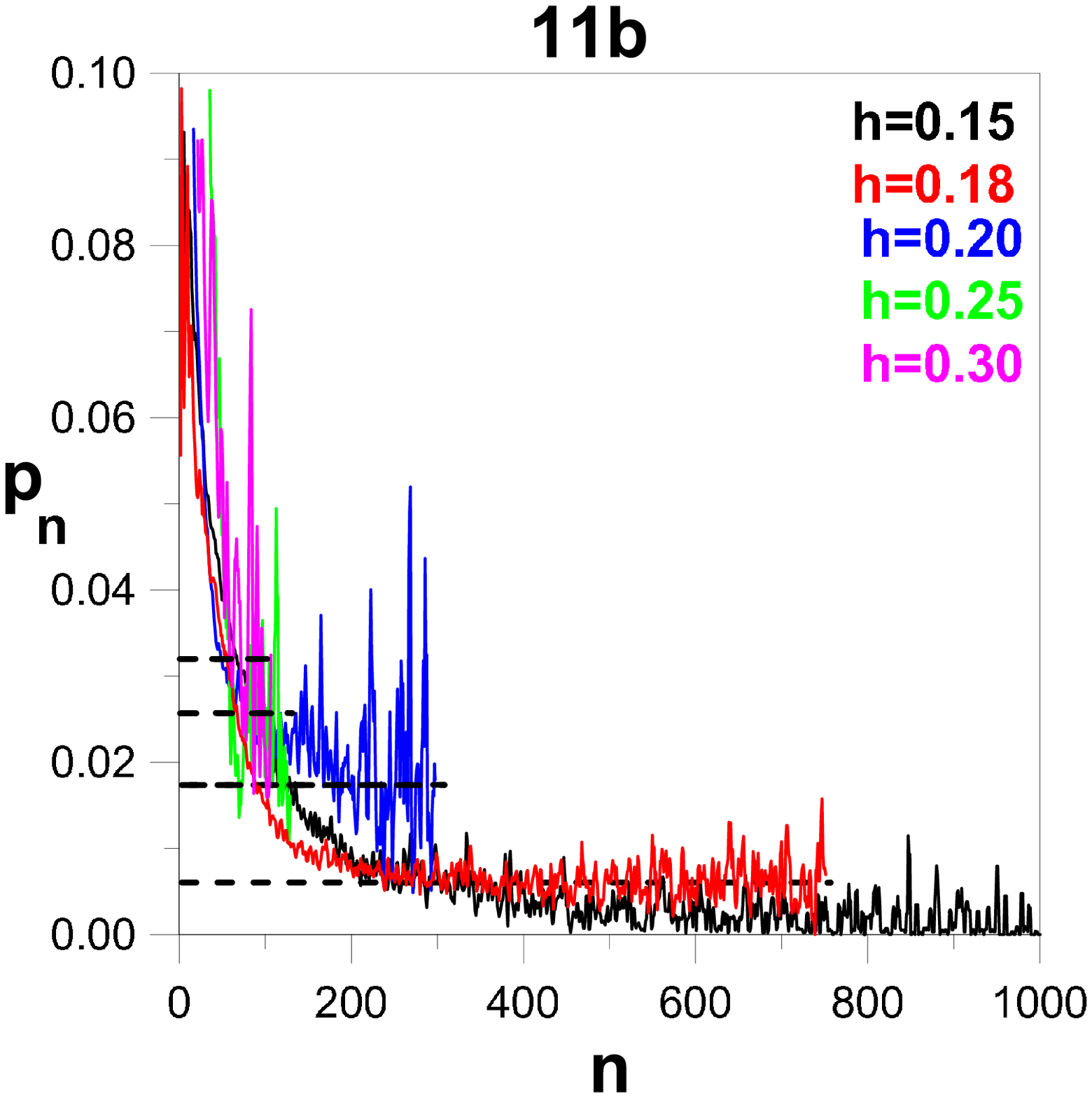}}
 \caption{
 (\ref{fig:11}a) The logarithmic proportions $log(\frac{dN_n}{N_o})$ of escapes
 for $h=0.126, 0.20, 0.25, 0.30, 0.35$ and $0.40$ as  functions of the number of
 intersections $n$.
 (\ref{fig:11}b) The probability $p_n=\frac{dN_n}{N_n}$ of escapes as a function
 of $n$ for various values of the energy $h$.}
\label{fig:11}
\end{figure}

We proceed now to a statistical analysis of the escapes. This is done
in four ways:
\renewcommand{\theenumi}{\alph{enumi}}
\renewcommand{\labelenumi}{(\theenumi)}

\begin{enumerate}
 \item First we separate the orbits into ordered and chaotic $+$ escaping. In
 Fig. \ref{fig:10}a we give the proportion of the chaotic orbits $N_{cha}/N_o$
 (including the escaping orbits) as a function of the energy $h$. Namely we
 populate the whole circle inside the limiting curve $x^2 +\dot{x}^2=2h$ by
 initial conditions of $N_0$ orbits (where $N_0=\pi (250)^2\simeq 200000$) and
 we find the total numbers of chaotic $+$ escaping orbits $N_{cha}$. Then the
 number of ordered orbits is $N_0-N_{cha}$. When $h$ is very small the
 proportion of ordered orbits is almost $100\%$. As $h$ increases the proportion
 of chaotic orbits increases and for $h=0.125$ this proportion is about $50\%$
 (Fig. \ref{fig:10}a). When $h$ goes beyond the escape perturbation $h=0.125$
 the proportion of chaotic orbits increases, but most chaotic orbits escape to
 infinity. Only a very small proportion of chaotic orbits near higher order
 unstable periodic orbits inside islands of stability do not escape (like the
 orbits near the unstable periodic orbit (b) of Fig. \ref{fig:06}).

 The proportion of chaotic orbits (including the escaping orbits) can be given
 by the approximate formula (Fig. \ref{fig:10}a)
 \begin{equation}
  N_{cha}/N_0=0.5[1+ ~\tanh{ (30.0h -4.0)]}.
 \end {equation}

 For small $h$ this proportion tends to zero, while for large $h$ it tends to 1.

 \item We calculate the proportion of direct escapes (Fig. \ref{fig:10}b),
 $dN_o/N_o$ (where $dN_o$ is the number of orbits that escape without any
 intersection with the axis $y=0$) as a function of the energy $h$. The direct
 escape regions are red in Figs. \ref{fig:05}, \ref{fig:08}, \ref{fig:09}a,b.
 This proportion is small for $h$ slightly larger than $h_{esc}= 0.125$ and
 increases fast with $h$, reaching more than $90\%$ for $h=0.7$.

 \item Then we study the escape time. The time is represented by the number of
 intersections $n$ before escape. Most escapes take a long time for $h$
 slightly above $h_{esc}$ and take place faster as $h$ increases. In Fig.
 \ref{fig:11}a we give the logarithm of the proportion of escaping orbits
 $d~N_n/N_0$ where  $d~N_n$ is the number of escapes after the $n$th
 intersection (upwards, $\dot{y}>0$) and before the $(n+1)$ intersection (i.e.
 there is no $(n+1)$ intersection). We see that the escape rates are decreasing
 with $n$. The escape rates $d~N_n/N_0$ are  large for small $n$ and decrease
 considerably for larger $n$. For $h\geq 0.35$ the escapes take place quite fast.

 \item Finally, we calculate the probability of escape as a function of  time
 represented by the number of intersections $n$ for various values of the
 energy. Namely the probability $p_n$ is the ratio $d~N_n/N_n$ where $N_n$
 is the number of orbits that have not yet escaped before the
 $n$th intersection
   \begin{equation} \label{func:EscProb}
    p_n =\frac {d~N_n}{N_n}.
   \end{equation}
This probability was calculated for the system (\ref{func:Ham}) and other
similar systems in previous papers \citep{Contop93,Siopis95a,Siopis95b,Siopis96}.
It was found that $p_n$ tends to a constant value $p$, independent of the
initial conditions of non-ordered orbits. For large times (large $n$) the
probability of escape tends to zero if the energy is smaller than a critical
value $h_{cr}$, larger than the escape energy $h_{esc}$. However, if $h$ is
larger than $h_{cr}$, the probability $p_n$ for large $n$, tends to a constant
value $p$ larger than zero. This quantity depends only on the energy $h$. In our
present notation the limiting probability $p$ is proportional to a power of the
quantity $(h-h_{cr})$, i.e.
  \begin{equation} \label{func:EscPL}
    p~\propto (h-h_{cr})^\alpha,
   \end{equation}
where $h_{cr}$ is the critical value of $h$, and the exponent $\alpha$ is
approximately $\alpha =0.5$. This value of $\alpha$ was found to be the same for
different dynamical systems \citep{Siopis95a,Siopis95b}.

In Fig. \ref{fig:11}b we give the value of $p_n$ as a function of $n$ for various
values of $h$. When $h= 0.15$ the probability $p_n$ tends to $p=0$. On the other
hand when $h=0.18, 0.20, 0.25, 0.30$ the probabilities tend to
$p=0.006, 0.018, 0.026, 0.034$. These values can be approximated by the formula
\begin{equation}
p=0.11(h - h_{cr})^{0.53}.
\end{equation}
where $h_{cr}=0.175$. This formula is consistent with the universal formula
found by \cite{Siopis95a,Siopis95b}.

The statistics of escapes in various dynamical systems has been a subject of
considerable interest in recent years. The main question is whether the various
chaotic phenomena (like Poincar\'{e} recurrences, correlations and escapes)
decay exponentially in time or according to power laws
\citep[e.g.][]{Crista08,Venege09}. The power laws are supposed to be connected
to the existence of stickiness in mixed dynamical systems that contain islands
of stability. However a detailed study of the effects of stickiness on the
escapes should be the study of a future research.

\end{enumerate}

\section{Periodic Orbits in the Manko-Novikov metric
(\ref{func:MNmetric})} \label{sec:PerOrbMN}

The Manko-Novikov (MN) spacetime depends on a parameter $q$ that measures the
quadrupole moment deviation of this metric from the Kerr metric with the same
mass $M$ and spin $S$. There have been proposals and attempts to constrain such
deviations from observational data (see e.g.
\cite{PsaJoh10,Bambi11,BamBar11a,BamBar11b} and references therein). The final
stages before the plunge (escape) in a non-Kerr spacetime like the MN have
certain astronomical interest (see \cite{BamBar11b}). From this point of view
our work aims to exploit the non-integrability of the MN metric in order to give
a detailed example of a such a final stage.

The geodesic orbits of a test particle of mass $\mu$ are described as equations
of motion of the following Lagrangian
\begin{equation}
L=\frac{1}{2}~\mu~g_{\mu\nu}~ \dot{x}^{\mu} \dot{x}^{\nu}.
\label{LagDef}
\end{equation}
The MN metric has two integrals of motion \citep{Gair08,Lukes10}, namely the
energy (per unit mass)
\begin{equation} \label{func:EnCon}
E=-\frac{\partial L}{\partial \dot{t}}/\mu=
f (\dot{\gamma} - \omega~ \dot{\varphi}),
\end{equation}
and the z-component of the angular momentum (per unit mass)
\begin{equation} \label{func:AnMomCon}
L_z =\frac{\partial L}{\partial \dot{\varphi}}/\mu=
f \omega (\dot{t} - \omega~ \dot{\varphi})+ f^{-1} \rho^2 \dot{\varphi},
\end{equation}
where the dots mean derivatives with respect to the proper time. The Kerr metric
has one more integral of motion, the so-called Carter constant \citep{Carter68},
thus it is an integrable system. However, the MN model is non-integrable
and it allows the appearance of chaos.

The motion on a meridian axis $(\varphi = cont)$ in the MN system satisfies the
relation
\begin{equation} \label{func:Veff}
\frac{1}{2} (\dot{\rho}^2 + \dot{z}^{2}) + V_{eff} (\rho, z)=0,
\end{equation}
where the effective potential $V_{eff} (\rho, z)$ depends on $q$, $E$ and
$L_z$. Thus the motion takes place inside the CZV
\begin{equation} \label{func:CZV}
 V_{eff} \equiv \frac{1}{2} e^{-2\gamma}\left[f-E^2+\left(\frac{f}{\rho}
 (L_z-\omega E) \right)^2\right] = 0.
\end{equation}
Studies of the orbits in the MN metric were made by
\cite{Gair08,Apostolatos09,Lukes10,Contop11}. In the present paper we study
on detail the periodic orbits, chaos and escapes in this system.

In the following we use the values $q=0.95$, $M=1$ and $\chi = S/M^2 = 0.9$
for the spin. We find the orbits for a fixed value of the z-angular momentum
$L_z =3$, and various values of $E$.

\begin{figure}[htp]
 \centerline{\includegraphics[width=0.5\textwidth] {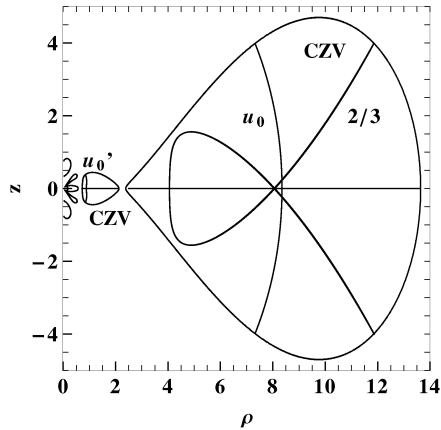}}
 \caption{
 Three stable periodic orbits  inside the outer  CZV: $u_0$, $2/3$ and $z=0$
 and two periodic orbits inside the inner CZV $u'_0$ (stable) and $z=0$
 (unstable). There are also 5 CZVs that reach the axis $\rho=0$. The line
 segment $-0.43589\leq z\leq 0.43589$ on the axis $\rho=0$ represents the
 horizon of the bumpy black hole. The horizon is broken at the point $\rho=z=0$.
}
 \label{FigOrPeInOutCZV}
\end{figure}

The form of the CZV in the case $E=0.95$ is given in Fig. \ref{FigOrPeInOutCZV}.
It consists of two main closed curves and 5 more curves of small extent reaching
the axis $\rho = 0$. The line segment $-k \leq z \leq k$ ($k= 0.43589$ for
$M=1,~\chi=0.9$), along the axis $\rho =0$, represents the horizon and any orbit
reaching this axis escapes into the bumpy black hole. All orbits inside the 5
curves that are close to the horizon escape into the bumpy black hole.

On the other hand, the orbits starting inside the main closed CZVs remain inside
these curves for ever. In particular there are two simple periodic orbits inside
the outer CZV, and two more simple periodic orbits inside the inner CZV. The
periodic orbits $u_0$ (outer) and $u'_0$ (inner) intersect perpendicularly the
$\rho$ - axis while the other two periodic orbits are straight lines along the
$z=0$ axis. Furthermore, we mark a stable resonant orbit $2/3$. The orbits
$u_0, u'_0$ and the outer $z=0$ orbit are stable, while the inner $z=0$ is
unstable.

\begin{figure}[htp]
 \centerline{\includegraphics[width=\textwidth] {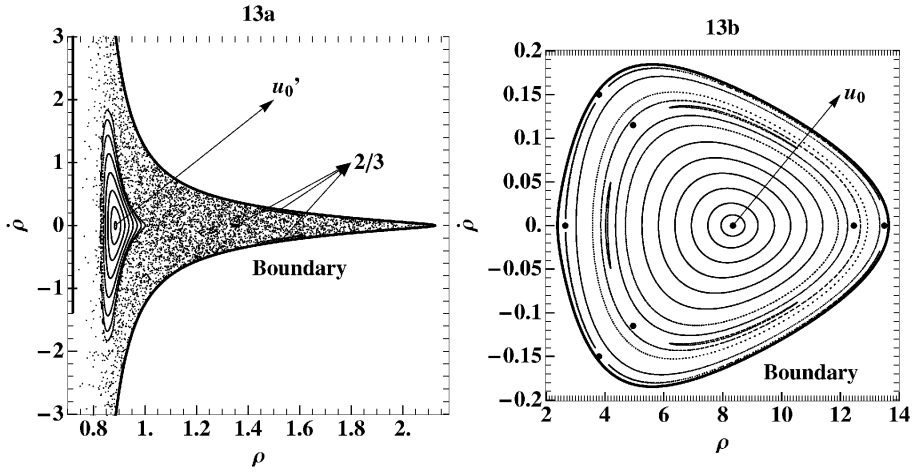}}
 \caption{
 Invariant curves and islands on a surface of section $(\rho,\dot{\rho})$
 $(z=0)$ for $E=0.95$. (\ref{FigPsInOut}a) Islands and chaos in the inner region.
 (\ref{FigPsInOut}b) Invariant curves and islands in the outer region. The
 unstable periodic orbits are marked with big dots.
}
 \label{FigPsInOut}
\end{figure}

If we take a surface of section $(\rho,\dot{\rho})$ the intersections
$(z=0,\dot{z}>0)$ of most orbits inside the outer CZV are along closed invariant
curves around the point $u_0$ that represents the periodic orbit $u_0$ (Fig.
\ref{FigPsInOut}b). However, there are also islands of stability around the
stable (resonant) periodic orbits, and some chaos between these islands, around
the unstable periodic orbits (Fig. \ref{FigPsInOut}b).

\begin{figure}[htp]
 \centerline{\includegraphics[width=0.5\textwidth] {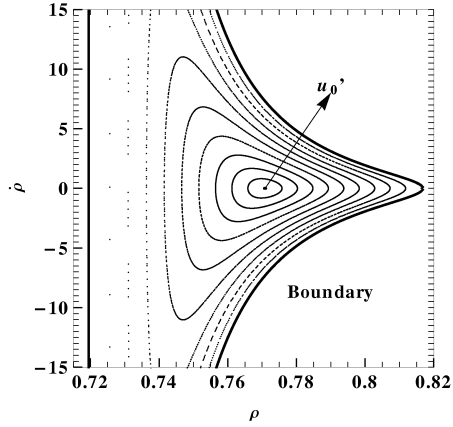}}
 \caption{
 Invariant curves around the periodic orbit $u'_0$ for $E=0.50$. The boundary
 represents the orbit $z=0$, which is stable in this case.
}
\label{FigPSE05}
\end{figure}

On the other hand, most orbits in the inner region (inside the inner CZV) are
chaotic (Fig. \ref{FigPsInOut}a) although there is a large island of stability
and three small islands of type $2/3$. The chaotic character of these orbits is
related to the instability of the orbit $z=0$, which is represented by the
boundary of the inner region on the surface of section. However, for a large
range of values of $E$ smaller than $E=0.67$ the orbit $z=0$ is stable and most
orbits around $u'_0$ are ordered (e.g. for $E=0.5$, Fig. \ref{FigPSE05}).

\begin{figure}[htp]
 \centerline{\includegraphics[width=\textwidth] {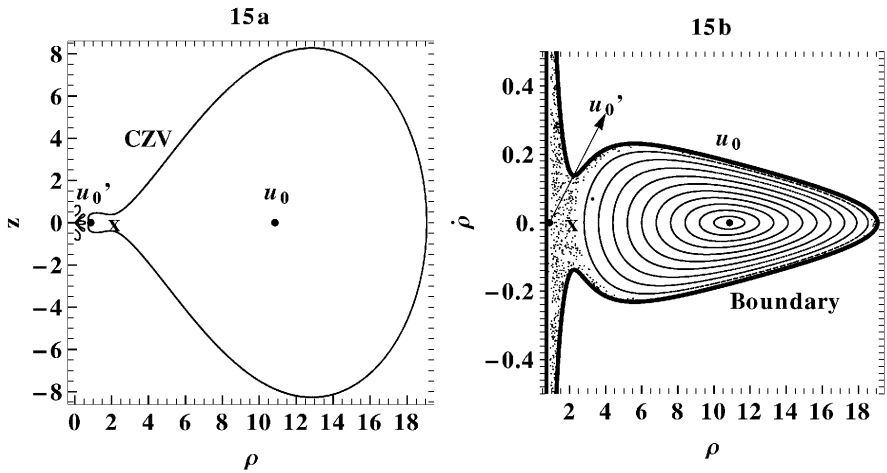}}
 \caption{
 (\ref{FigE096}a) The two main CZV are joined into one for $E=0.96$. The point
 \texttt{x} represents an unstable orbit that is generated at the saddle point
 when the two curves join. (\ref{FigE096}b) The corresponding surface of section
 $(\rho,\dot{\rho})$. In this case the boundary, which represents the orbit
 $z=0$, is unstable. Chaos is dominant in the inner region and close to the
 boundary.
}
\label{FigE096}
\end{figure}

If $E$ increases beyond $E=0.9504$ the inner and outer CZVs are joined into a
common CZV (Fig. \ref{FigE096}a for $E=0.96$). When the two curves join, an
unstable periodic orbit is formed, which intersects perpendicularly the $z=0$
axis (point x in Fig. \ref{FigE096}a) and exists for larger values of E. At the
same time the two periodic orbits $z=0$ join into one unstable orbit. The inner
and outer regions are also joined on the surface of section $(\rho, \dot{\rho})$
(Fig. \ref{FigE096}b). The periodic orbit $z=0$ is represented now by the common
boundary of the two regions and close to it there is some chaos extending all
the way around the orbit $u_0$. Chaos is also dominant inside (i.e. on the left
of) the orbit x (Fig. \ref{FigE096}b).

\begin{figure}[htp]
 \centerline{\includegraphics[width=0.5\textwidth] {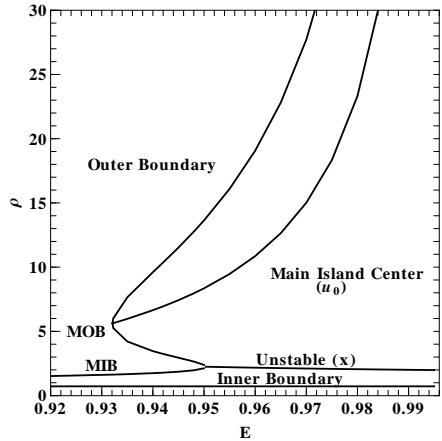}}
 \caption{
  Characteristics of the periodic orbits $(u_0,x)$ and the boundaries of the
  main permissible regions ($\rho$ is given as a function of the energy E).
}
\label{FigEpMfLE}
\end{figure}

Figure \ref{FigEpMfLE} gives the characteristics of the periodic orbits $u_0$
and $x$ and the boundaries of the permitted motions for $z=\dot{\rho}=0$. The
inner boundary is at an almost constant $\rho\simeq 0.72$. The outer boundary
increases considerably with increasing $E$ and tends to infinity as
$E\rightarrow 1$. When $E>1$ the orbits escape to infinity. When $E<0.9504$ the
permissible region splits into two and we have a middle outer boundary (MOB)
(Fig. \ref{FigEpMfLE}) which is the inner boundary of the outer region, and a
middle inner boundary (MIB) which is the outer boundary of the inner region.

As $E$ decreases the outer region shrinks, and disappears for $E\simeq 0.9321$.
Then the orbit $u_0$ also disappears after becoming  just one point. Then the
orbit in 3 dimensions is a circle on the equatorial axis around the center
$\rho =0$. For smaller $E$ only the inner closed CZV persists.

\begin{figure}[htp]
 \centerline{\includegraphics[width=\textwidth] {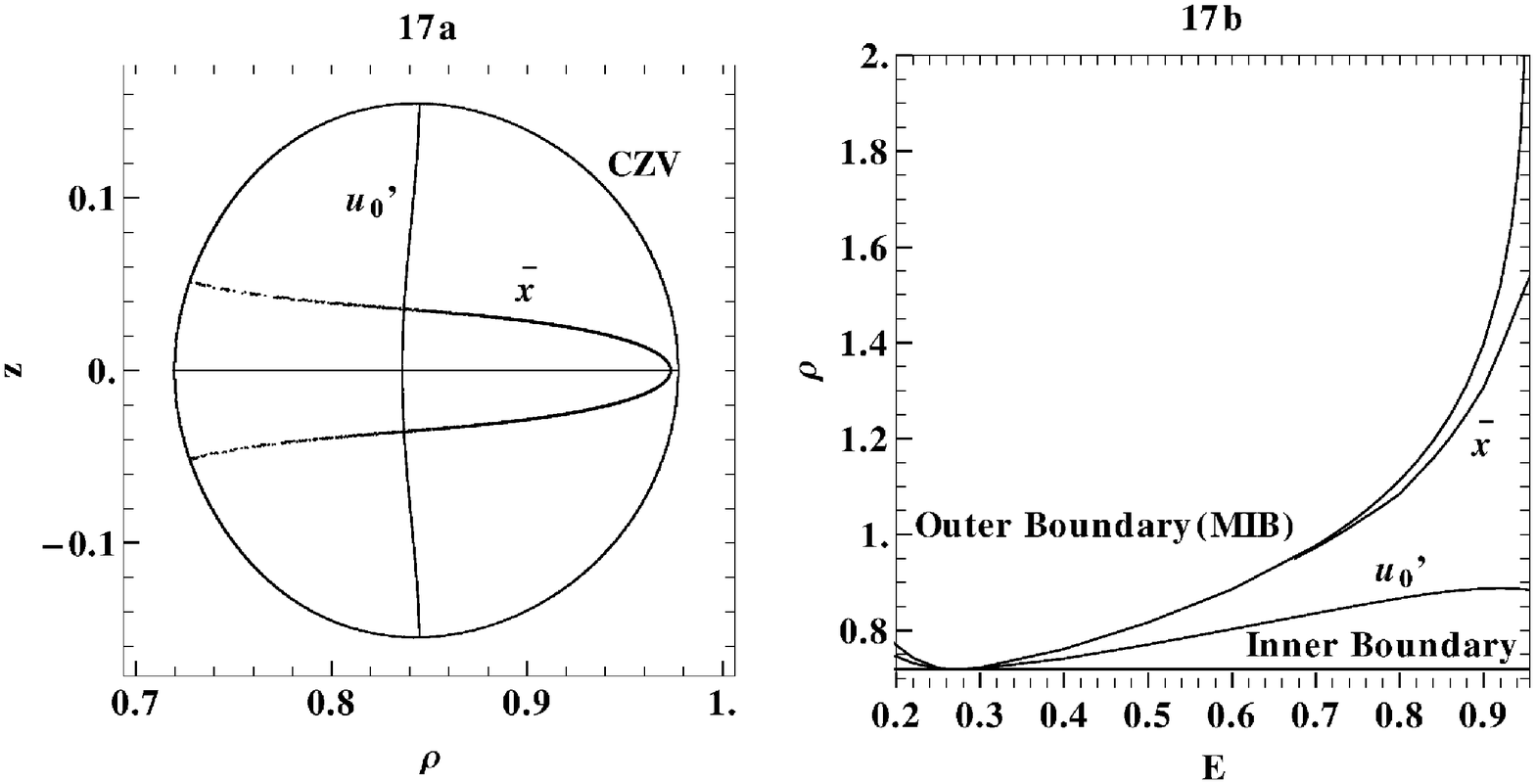}}
 \caption{
 (\ref{FigInReCh}a) Periodic orbits $u'_0$ and $\bar{x}$ for $E=0.7$. The family
 $\bar{x}$ was generated at the transition of the orbit $z=0$ from stability to
 instability as $E$ increases. (\ref{FigInReCh}b) The characteristics of the
 families $u'_0,\bar{x}$ and the boundaries of the (inner) permissible region.
 }
\label{FigInReCh}
\end{figure}

The position of the  orbit $u'_0$ is close to the inner boundary, and cannot be
distinguished from the inner boundary in the scale of Fig. \ref{FigEpMfLE}.
In fact, the distance of this orbit from the inner boundary decreases, as $E$
decreases, and goes to zero at about $E=0.28$ (Fig. \ref{FigInReCh}b). For the
same value of $E$ the outer boundary of the inner region (MIB) reaches the inner
boundary. Then the orbit $u'_0$ in 3 dimensions is circular on the equatorial
plane.

As we have mentioned above the orbit $z=0$ of the inner region becomes stable
when $E$ decreases below $E\simeq 0.67$. As $E$ increases above this critical
value, a stable family $\bar{x}$ that crosses perpendicularly the $z=0$ axis
bifurcates from $z=0$ (Fig. \ref{FigInReCh}a,b). As $E$ increases further this
family becomes unstable,  generating by bifurcation higher order periodic orbits.
E.g. for $E=0.95$ one can see 3 small islands around this unstable family (on
the right hand side of Fig. \ref{FigPsInOut}a).

\begin{figure}[htp]
 \centerline{\includegraphics[width=0.7\textwidth] {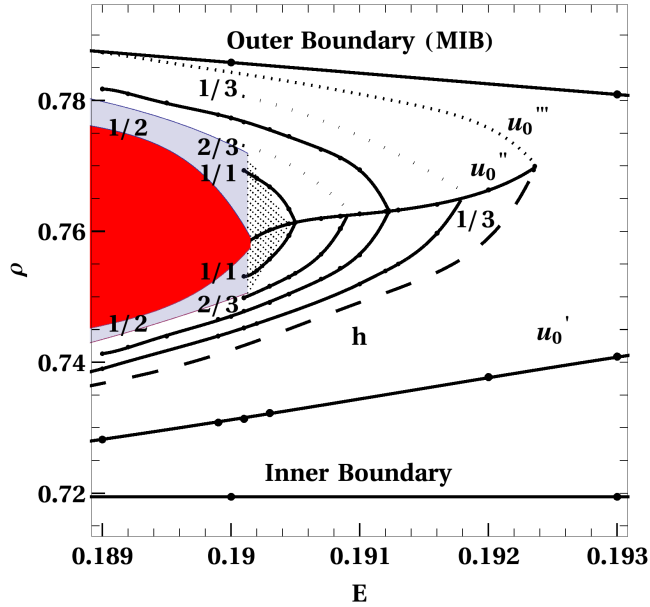}}
 \caption{
 Characteristics of the families $u'_0, u^{''}_0, u^{'''}_0$ and of some
 resonant bifurcations from the family $u''_0$ for relatively small values of
 the energy E. The dotted region indicates chaos. The dashed line gives the
 position of the first homoclinic point of the orbit $u^{'''}_0$, and the gray
 region indicates escaping orbits, while the red region gives the initial
 conditions (for $z=0$) of orbits that escape directly.
 }
\label{FigEpMfs}
\end{figure}

When E decreases below $E=0.28$ the inner CZV increases again (left side of Fig.
\ref{FigInReCh}b), and the orbit $u'_0$ deviates from the inner boundary. But,
for $E<0.198$ the distance of $u'_0$ from the inner boundary decreases again
(Fig. \ref{FigEpMfs}), although it does not tend to zero.

The most strange evolution occurs when E becomes smaller than $E=0.19236$. Then
a couple of simple periodic orbits crossing perpendicularly the $z=0$ axis is
formed out of nothing. One orbit $(u^{''}_0)$ is stable, and the other
$(u^{'''}_0)$ unstable (Fig. \ref{FigEpMfs}). This is called a "tangent
bifurcation" \citep{Contop02}.

\begin{figure}[htp]
 \centerline{\includegraphics[width=\textwidth] {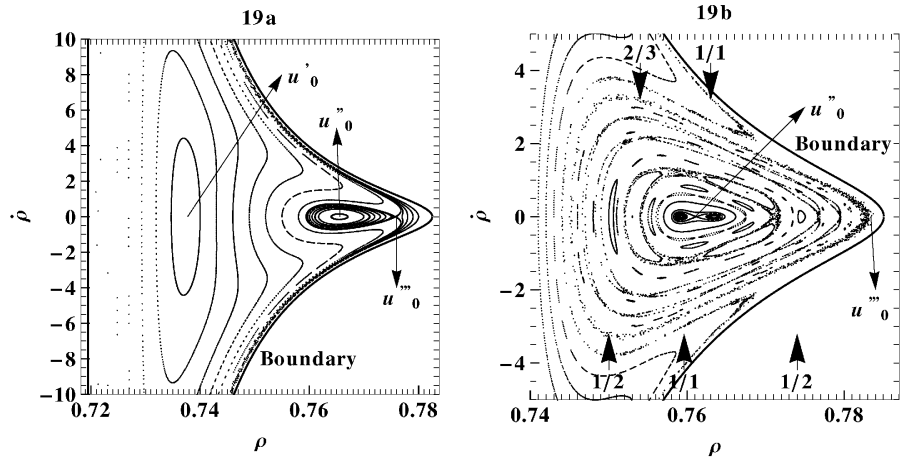}}
 \caption{
  (\ref{FigSSTB}a) A surface of section  $(\rho,\dot{\rho})$ for $E=0.192$ (just
  below the value of $E=0.19236$ when the orbits $u^{''}_0, u^{'''}_0$ are
  formed at a tangent bifurcation). Invariant curves surround the orbits
  $u'_0, u^{''}_0$ and are also near the boundary. A little chaos appear near
  the unstable orbit $u^{'''}_0$ and its asymptotic curves. (\ref{FigSSTB}b) A
  surface of section for $E=0.19045$. It contains several islands around
  periodic orbits that have bifurcated from $u^{''}_0$, like $1/2,~2/3$ and
  $1/1$. Several chaotic regions appear between the islands, but they are
  separated from each other by invariant curves around $u^{''}_0$. For this
  value of E the orbit  $u^{''}_0$ is unstable, while the boundary (representing
  the orbit $z=0$) is stable.
 }
\label{FigSSTB}
\end{figure}

From the unstable point $u^{'''}_0$ start two stable and two unstable asymptotic
manifolds (Fig. \ref{FigSSTB}a). The inner stable and unstable asymptotic
manifolds surround the orbit $u^{''}_0$ and intersect at an infinite number of
homoclinic points. The first homoclinic point is on the axis $\dot{\rho} =0$, on
the left of the orbit $u^{''}_0$. The other couple of stable and unstable
asymptotic curves approach the boundary and surround the orbit $u'_0$ on the
left of the figure.

Near the unstable orbit $u^{'''}_0$ and its homoclinic points there is some
chaos (Figs. \ref{FigEpMfs} and \ref{FigSSTB}a). As $E$ decreases several
periodic orbits bifurcate from $u^{''}_0$ and near all the unstable orbits there
are chaotic regions. These chaotic regions increase as $E$ decreases. Two
examples of bifurcating families, namely $2/3$ and $1/3$ are shown in Fig.
\ref{FigEpMfs}. The orbit $u^{''}_0$ remains stable from its generation at
$E=0.19236$ down to $E=0.1912$. At $E=0.1912$ the orbit $u^{''}_0$ becomes
unstable and a double period $1/2$ stable orbit bifurcates there towards smaller
values of $E$. In Fig. \ref{FigSSTB}b this orbit has receded from $u^{''}_0$ and
two islands of stability are formed, one around each point. The orbits of the
islands go alternatively from one island to the other.

The orbit $u^{''}_0$ remains unstable for a small interval $\Delta E$ and then
it becomes again stable, generating an unstable periodic orbit $1/2$.

As E comes close to $E=0.194045$ the family $u^{''}_0$ becomes again unstable
for a small interval $\Delta E$. At the transition to instability it generates
two different stable periodic orbits of equal period $1/1$. In Fig.
\ref{FigSSTB}b the orbit $u^{''}_0$ is unstable and the two islands around it
refer to different orbits. For a little smaller $E$ the orbit $u^{''}_0$ becomes
again stable and generates two unstable  periodic orbits $1/1$.

As $E$ decreases further there is an infinity of transitions to instability and
stability of orbits of double period $1/2$ and of equal period $1/1$. At the
limit of this infinite sequence of transitions  the orbit $u^{''}_0$ reaches the
escape region (Fig. \ref{FigEpMfs}) and does not exist for smaller values of $E$.

This phenomenon of infinite transitions to instability and stability along the
same family was found also in other dynamical systems, e.g. the system given by
Eq. (\ref{func:Ham}) of the present paper \citep{Contop80}.

\begin{figure}[htp]
 \centerline{\includegraphics[width=\textwidth] {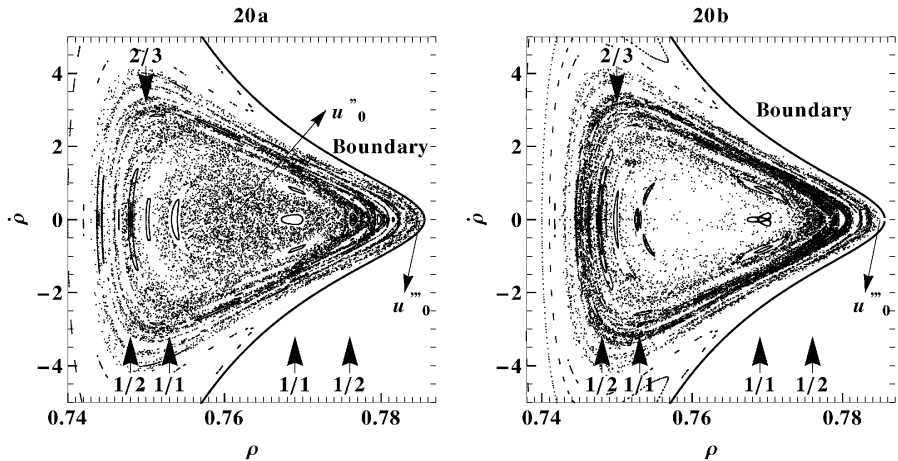}}
 \caption{
  (\ref{FigSSEs}a) A surface of section $(\rho,\dot{\rho})$ for $E=0.19015$. In
  this case most of the chaotic regions have joined. However, there are also
  islands, like $1/2,~2/3,~1/1$.   (\ref{FigSSEs}b) A surface of section for
  $E=0.1901$ below the escape value of $E=E_{esc}$ where escapes start to appear.
  The chaotic region of Fig. \ref{FigSSEs}a is now a region of escapes, but we
  see also the same islands.
}
\label{FigSSEs}
\end{figure}

The chaotic regions of Fig. \ref{FigSSTB}b are  separated from each other by
invariant curves closing  around $u^{''}_0$. For a little smaller $E$ most of
these invariant curves are destroyed, as $E$ decreases, and a large chaotic
region is formed (Figs. \ref{FigEpMfs} and \ref{FigSSEs}a) for $E=0.19015$. This
large chaotic region contains several islands of stability, like $1/1,~1/2,~2/3$
etc (Fig. \ref{FigSSEs}a), that recede from $u^{''}_0$ as $E$ decreases (see
the characteristics of the corresponding periodic orbits in Fig. \ref{FigEpMfs}).

\section{Escapes from the Manko-Novikov system
 (\ref{func:MNmetric})}\label{sec:EscMN}

In the MN system (including the Kerr system) there are two types of escapes of
orbits: (a) Escapes to infinity when $E \gtrsim 1$, and (b) escapes to the
central bumpy black hole for small values of E.

The first type of escapes (escapes to infinity) is similar to the escapes in the
Keplerian two-body problem (escapes along hyperbolic orbits). The permissible
region increases in size as the energy increases. This size tends to infinity as
the energy approaches the energy of the parabolic motions.

On the other hand, the escapes to the bumpy black hole occur when the CZVs open
inwards as the energy decreases and allow motions towards the bumpy black hole.

\begin{figure}[htp]
 \centerline{\includegraphics[width=\textwidth] {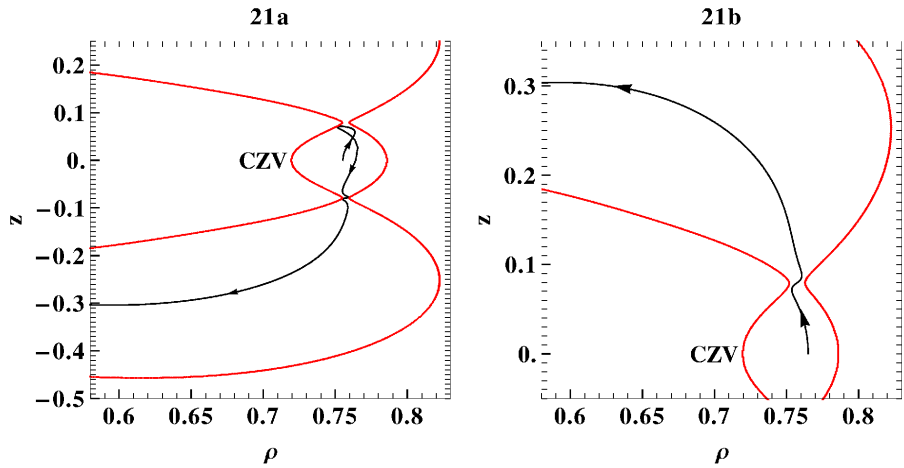}}
 \caption{
 Escapes in the MN metric Fig. (\ref{FigEscOr}a) downwards $(E=0.1901)$ and
 Fig. (\ref{FigEscOr}b) upwards $(E=0.1899)$.
 }
\label{FigEscOr}
\end{figure}

In the MN case, that we study here, when $E=0.1901$ the CZV of the inner region
(the outer region does not exist) joins two CZVs, above and below it and orbits
from the inner region are allowed to escape to the bumpy black hole $(\rho =0)$.
The aforementioned  CZVs above and below the central region are the second and
fourth leaf-like CZVs respectively (counting clockwise) in Fig.
\ref{FigOrPeInOutCZV} from the set of five CZVs touching the broken horizon of
the bumpy black hole $(\rho=0)$. These two CZVs expand outwards on both sides of
the axis $\dot{\rho}=0$, while $E$ decreases, until (for $E\approx 0.1901$) they
reach the expanding central inner region (Fig. \ref{FigEpMfs}), which till then
contains only bounded orbits. After the connection, orbits of the inner region
may escape downwards (Fig. \ref{FigEscOr}a) or upwards (Fig. \ref{FigEscOr}b) to
the central bumpy black hole.

In the MN case the escaping orbits are chaotic orbits that undergo chaotic
scattering. In fact, if we compare Fig. \ref{FigSSEs}a $(E=0.19015)$ where no
escapes are permitted, and Fig. \ref{FigSSEs}b where we have many escapes we see
a great similarity. That is the large chaotic domain around $u^{''}_0$ of
Fig. \ref{FigSSEs}a is transformed into an escape domain in Fig. \ref{FigSSEs}b.
We see several islands in Fig. \ref{FigSSEs}b that are slightly changed from the
islands of Fig. \ref{FigSSEs}a (the main change is that the islands $1/1$ in
Fig. \ref{FigSSEs}a are surrounded by 3 secondary islands in Fig.
\ref{FigSSEs}b).

This similarity explains why the chaotic domain of Fig. \ref{FigEpMfs} (around
$u^{''}_0$) is transformed into an escape domain for $E\leq 0.1901$. In
particular the periodic orbit $u^{''}_0$ of Fig. \ref{FigSSEs}a reaches the
point where the main CZV of the inner region joins the upper and lower CZVs and
for smaller $E$ this orbit escapes to the bumpy black hole, i.e. the periodic
orbit does not exist any more.

\begin{figure}[htp]
 \centerline{\includegraphics[width=0.5\textwidth] {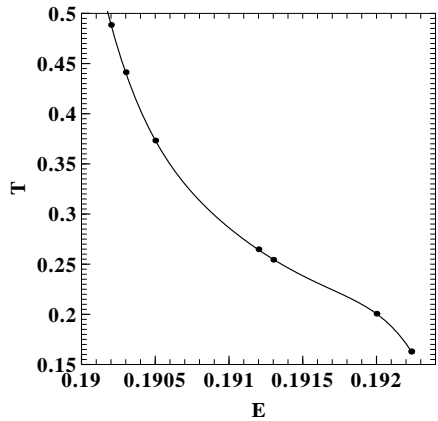}}
 \caption{
 The period of the orbit $u'_0$ as a function of the energy $E$.
 }
\label{FigEnPeIn}
\end{figure}

This case is very similar to the cases $y = \pm \surd 2x$ (orbits $(a)$ and
$(a')$) of the system (\ref{func:Ham}). There are two more similarities between
the two cases. First is the sequence of infinite transitions to instability and
stability of the orbits $u^{'}_0$ and the orbits $(a)$ and $(a')$. The second
similarity is in the periods of the orbit $u^{''}_0$ and of the orbits $(a)$ and
$(a')$, which both tend to infinity as the orbits approach the termination point.
The period of the orbits $u^{''}_0$ is given in Fig. \ref{FigEnPeIn}, which
increases considerably as $E$ tends to $E=0.1901$ and presumably it tends to
infinity. The period of the orbits $y = \pm \surd 2x$ was also found to tend to
infinity \citep{Contop80}.

This behavior is consistent  with the Str\"{o}mgren  termination principle of
the families of periodic orbits \citep{Szebehely67}. According to this principle
the families either join other families, or terminate when their size, or their
energy, or their period becomes infinite.

\begin{figure}[htp]
 \centerline{\includegraphics[width=0.5\textwidth]
 {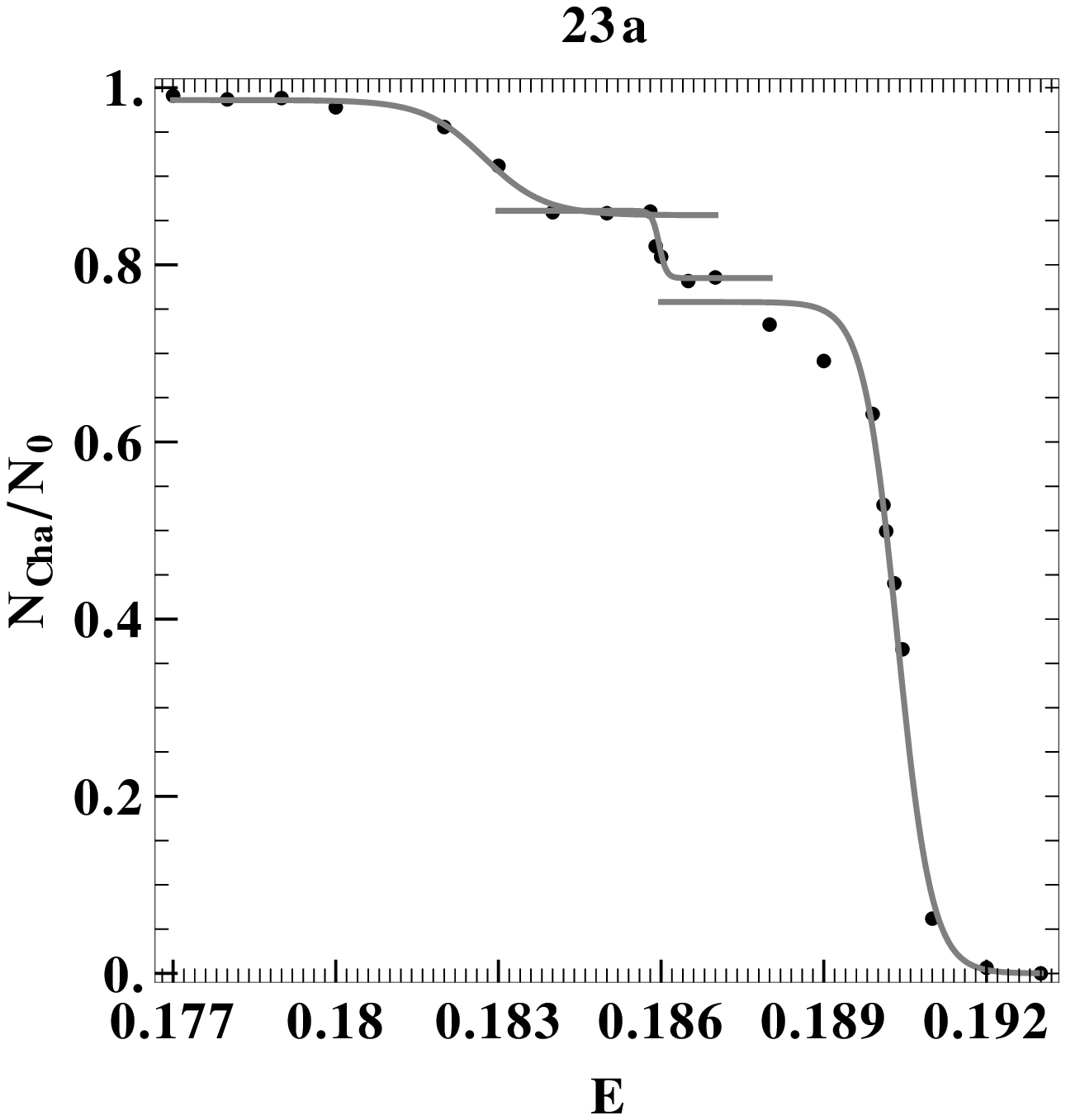}\includegraphics[width=0.5\textwidth] {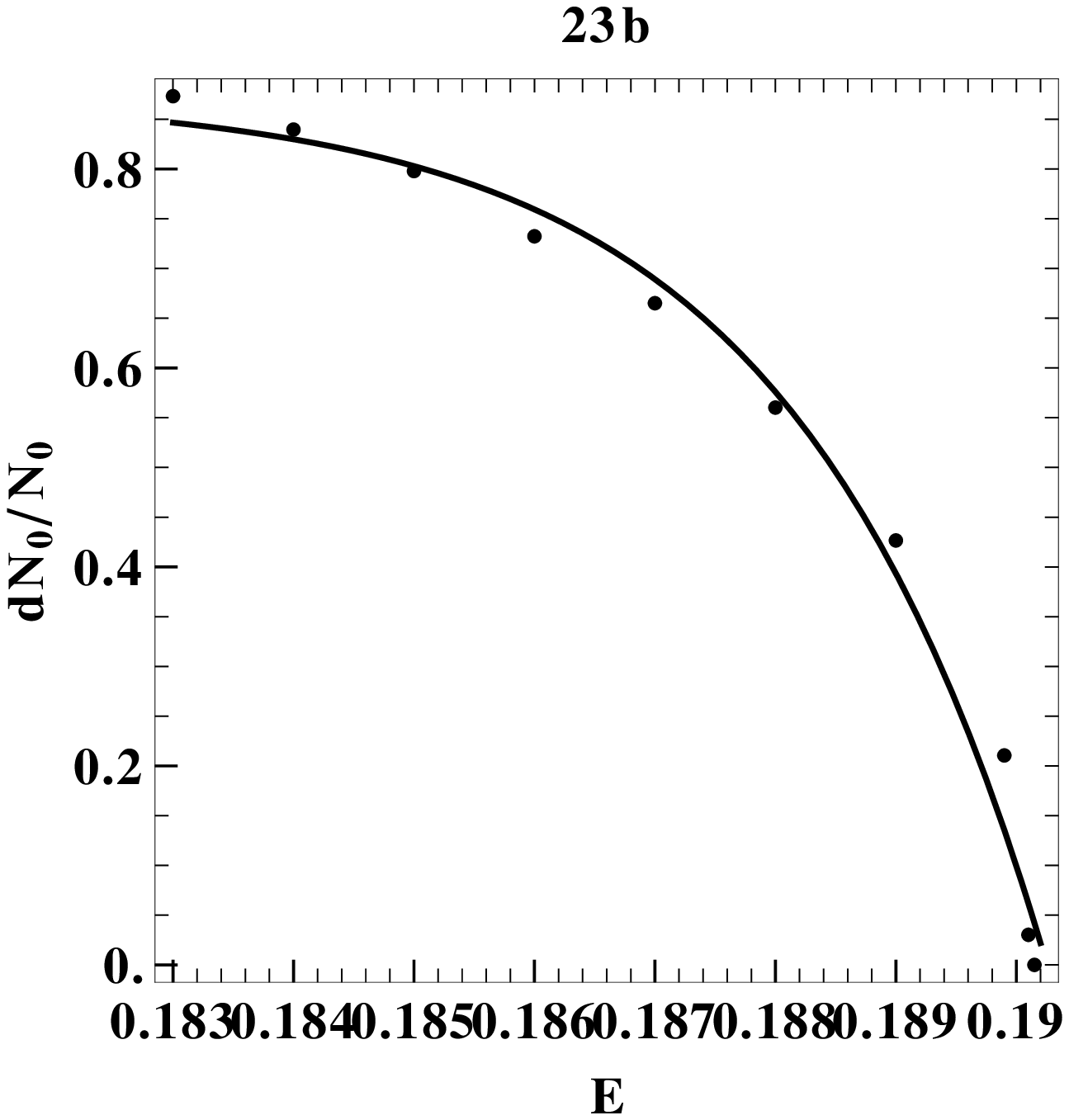}}
 \caption{
 (\ref{FigChaReP}a) The proportion of the chaotic orbits (including the escape
 orbits), with initial conditions along the axis $x$, as a function of the
 energy $E$.  (\ref{FigChaReP}b) The proportion $d N_0/N_0$ of direct escapes as
 a function of $E$.
 }
\label{FigChaReP}
\end{figure}

The proportion of chaotic $+$ escaping orbits $N_{cha}/N_0$ increases as the
energy decreases (Fig. \ref{FigChaReP}a). We calculate the proportion
$N_{cha}/N_0$ by taking initial conditions of orbits along the $x$-axis. This
proportion tends to zero for relatively large $E$ but it increases abruptly as
$E$ goes beyond $E_{esc}$ (to smaller $E$). An approximate formula for this
proportion is
\begin{equation}\label{func:approx1}
N_{cha}/N_o = 0.379[1+ \tanh (304-1597 E)]
\end{equation}
and can be applied in the interval $0.188<E<0.192$ (Fig. 25).
The curve (\ref{func:approx1}) is very similar to the corresponding curve of
the Hamiltonian (\ref{func:Ham}) (Fig. \ref{fig:10}a). However, for
$E \leq 0.186$ the proportion of escaping orbits increases further and tends to
$100\%$. In Fig. \ref{FigChaReP}a two more abrupt increases can be seen.
The first of these two increases takes place near $E \approx 0.186$, and around
it one can use an approximate formula
$N_{cha}/N_o = 0.823+0.038 \tanh (1399-7253 E)$. The second increase takes
place near $E \approx 0.183$ and the corresponding approximate formula is
$N_{cha}/N_o = 0.921 + 0.065 \tanh (163 - 892 E)$. This formula gives values
near $N_{cha}/N_o = 0.99$ for $E<$0.180.

When $E<E_{esc}$ most chaotic orbits escape to the bumpy black hole. However,
only a proportion $dN_0/N_0$ escape directly from the system without any further
intersection with the axis $z=0$. This proportion increases as the energy $E$
decreases (Fig. \ref{FigChaReP}b). The curve $dN_0/N_0$ of Fig.
\ref{FigChaReP}b is very similar to the corresponding curve of Fig.
\ref{fig:10}b. In Fig. \ref{FigEpMfs} we mark in red the initial conditions of
the directly escaping orbits. Most of the other chaotic orbits escape after one
or more intersections beyond the original point, with the $z=0$ axis. But, as we
see in Fig. \ref{FigEpMfs}, for $E<E_{esc}$ there are also orbits that do not
escape. In particular the periodic orbit $u'_0$ and orbits close to it do not
escape for much smaller values of $E$. Similarly the orbits that bifurcated from
$u^{''}_0$, like the orbits $1/3,~1/2,~2/3,~1/1$ (Fig. \ref{FigEpMfs}) exist
also for $E<E_{esc}$. However, all these families become unstable for small $E$.
In any case the existence of an infinity of families of unstable periodic orbits
is a similar phenomenon as the one observed in the system (\ref{func:Ham}).

\section{Conclusions} \label{sec:concl}

We considered the periodic orbits and escapes in two quite different dynamical
systems in order to find their common features. The first system is a system of
two coupled oscillators, while the second system is a perturbation of the Kerr
metric in General Relativity.

In the system (\ref{func:Ham}) escapes occur when the Curves of Zero Velocity
open and many orbits escape to infinity. In the second case we have escapes to
infinity, but we emphasized the escapes to the central bumpy black hole. In both
cases when the energy changes and reaches the escape value, there is a periodic
orbit that approaches the point where the CZV opens and its period tends to
infinity. Thus, the family of periodic orbits terminates at the escape energy.
Before the escape energy this family undergoes an infinity of transitions from
stability to instability and vice-versa, and at every transition there is a
bifurcation of an equal or double period family of periodic orbits, that does
not escape as the energy varies.

It seems that, in general, in systems with escapes there is a family of periodic
orbits that terminates at infinite period when the CZVs open.

Another common feature is that most chaotic orbits become escape orbits beyond
the escape energy (beyond means larger energy in the first system, but smaller
energy in the second system). However, for energies slightly beyond the escape
energy most orbits take  a rather long time to escape. As the energy increases
(decreases) the proportion of fast escapes increases considerably. The
proportion of chaotic ($+$ escaping) orbits increases abruptly as the energy
goes beyond the escape energy.

We studied the structure of the phase space and we distinguished the ordered and
chaotic domains. The ordered domains surround the positions of the stable
periodic orbits. Thus we found these periodic orbits and their bifurcations as
the energy varies. On the other hand, the chaotic domains surround the
asymptotic curves from the unstable periodic orbits.

As the two systems under study are quite different we studied them in detail
separately.

In the first case (of two coupled oscillations) we found the characteristics of
the periodic orbits and their stability. Then we found the structure of the
phase space on a surface of section for various values of the energy. As the
energy increases the chaotic domains increase, while all stable periodic orbits
become unstable. (However, new stable periodic orbits appear for larger energies
at tangent bifurcations.)

The chaotic domains are covered by the asymptotic curves of a main unstable
periodic orbit. As the energy goes beyond the escape energy most chaotic orbits
escape, either directly, or after a small or a large number of intersections
with the $y$=0 axis. Some asymptotic curves are split into successive pieces,
each piece making infinite rotations around an escape domain.

The statistics of chaos and escape are given by different graphs: (a) (b) the
proportion of chaotic and escaping orbits and the proportion of the directly
escaping orbits as functions of the energy, (c) (d) the proportion of the
escaping orbits at the $n$th intersection with respect to the initial total
number of orbits and with respect to the remaining orbits, as functions of the
number of iterations (which represents time).

In the system (\ref{func:MNmetric}) (Manko-Novikov metric) we found various
forms of the CZVs for various values of the energy, and the corresponding simple
periodic orbits. We found also surfaces of section with mostly ordered orbits,
for relatively large energies $E$ (but $E<$1), or a mixture of ordered and
chaotic orbits for smaller energies. We found the characteristics of the main
families of periodic orbits and their bifurcations.

We emphasized the appearance of a couple of periodic orbits (one stable, and one
unstable) at a tangent bifurcation, and their evolution. As the energy decreases
the stable family undergoes an infinity of bifurcations (at transitions from
stability to instability and vice versa). As the energy goes below a critical
value the period of this orbit goes to infinity and then this orbit escapes to
the bumpy black hole and this family of periodic orbits terminates.

On the other hand, the bifurcating families exist and do not escape, but become
unstable. We found the proportion of the chaotic regions as the energy decreases.
Beyond the critical energy most of the chaotic region contains escape orbits.
The proportion of the chaotic$+$escaping orbits increases abruptly below the
escape energy and tends to $100\%$ for smaller energies.

It is remarkable that these quite different systems have very similar forms of
the proportion of chaotic orbits (Figs. \ref{fig:10}a and \ref{FigChaReP}a) and
of the proportion of directly escaping orbits (Figs. \ref{fig:10}b and
\ref{FigChaReP}b). The relations of the periodic orbits with the chaotic and
escaping orbits are also quite similar. These similarities indicate that the
properties of periodic, ordered, chaotic and escaping orbits that we studied in
the present paper are very similar in generic dynamical systems.

\begin{acknowledgements}
 G. Lukes-Gerakopoulos was supported in part by the Research Committee of the
 Academy of Athens and by the DFG grant SFB/Transregio 7.
\end{acknowledgements}

\end {document}